\documentclass[12pt,a4paper]{article}
\usepackage[english]{babel}
\usepackage{graphicx}

\def\alt{\stackrel{<}{\sim}}
\def\agt{\stackrel{>}{\sim}}
\begin{document}
\setcounter{footnote}{0}
\setcounter{equation}{0}
\setcounter{figure}{0}
\setcounter{table}{0}
\vspace*{5mm}

\begin{center}
{\large\bf Symmetry Theory of the Anderson Transition }

\vspace{4mm}
I. M. Suslov \\
P.L.Kapitza Institute for Physical Problems, \\
117337 Moscow, Russia
\vspace{6mm}

\begin{minipage}{155mm}
{\rm
We prove the Vollhardt--W${\rm {\ddot o}}$lfle hypothesis that
the irreducible vertex $U_{{\bf kk}^\prime} (q)$ appearing in
the Bethe--Salpeter equation contains a diffusion pole (with the
observable diffusion coefficient $D (\omega, q)$) in the limit ${\bf k} +
{\bf k}^\prime \rightarrow 0$. In the quantum kinetic equation, the
quantity $U_{{\bf kk}^\prime} ({\bf q})$ plays the role of a transition
probability $W_{{\bf kk}^\prime}$ and its anomalous growth as
$D (\omega, q)\to 0$ is the physical reason for localization.
As $\omega \rightarrow
0$, the relation $D (\omega, q) = (-i \omega) d (q)$
holds in the localized phase, where $d(q)$ is a regular function of $q^2$,
related with the
properties of a typical wave function. The presence of a diffusion pole in
$U_{{\bf kk}^\prime} ({\bf q})$ makes it possible to represent the quantum
"collision operator" $\hat L$ as a sum of a singular operator ${\hat
L}_{sing}$, which has an infinite number of zero modes, and a regular
operator ${\hat L}_{reg}$ of a general form. Investigation of the response
of the system to a change in ${\hat L}_{reg}$ leads to a self-consistency
equation, which replaces the rough Vollhardt--W${\rm {\ddot o}}$lfle
equation. Its solution shows that $D (0, q)$ vanishes at the transition
point simultaneously for all $q$. The spatial dispersion of $D (\omega,
q)$ at $\omega \to 0$ is found to be $\sim 1$ in relative units. It is
determined by the atomic scale, and it has no manifestations on the scale
$q \sim \xi^{-1}$ associated with the correlation length $\xi$. The values
obtained for the critical exponent $s$ of the conductivity and the
critical exponent $\nu$ of the localization length in a $d$-dimensional
space, $s = 1\,\, (d > 2)$ and $\nu = (d - 2)^{-1} \,\,(2 < d < 4),\quad \nu
= 1/2 \,\,(d > 4)$, agree with all reliably established results. With respect
to the character of the change in the symmetry, the Anderson transition is
found to be similar to the Curie point of an isotropic ferromagnet with an
infinite number of components. For such a magnet, the critical exponents are
known exactly and they agree with the exponents indicated above. This
suggests that the symmetry of the critical point has been established
correctly and that the exponents have been determined exactly. }
\end{minipage} \end{center}

\newpage
\vspace{6mm}
\begin{center}
{\bf 1. INTRODUCTION} \\
\end{center}

It is now widely acknowledged (see, for example, Ref. 1, p. 76) that the
theory of phase transitions should, in principle, be constructed as a
symmetry theory. Specifically, the effective Hamiltonian of the system is
represented in the form
$$
H = H_c + \tau H_{int}\eqno(1)
$$
where $H_c$ is the critical-point Hamiltonian, possessing a high symmetry;
$H_{int}$ is a general operator which is compatible with the symmetry of
the total Hamiltonian $H$; and, $\tau$ is a parameter that measures the
distance to the transition. The most general motivation for the separation
(1) is that the set of Hamiltonians $H_c$ (for example, Hamiltonians of
different ferromagnets at the Curie point) should be separated from the
set of all Hamiltonians $H$ by imposing some kind of additional conditions
which can be interpreted as generalized symmetry requirements.

In this approach, the problem consists of determining the complete symmetry
of the Hamiltonian $H_c$; thus far, it has been impossible to do for
most phase transitions. For example, the well-known Landau  theory [2]
starts from the obvious symmetry of the Hamiltonian and does not take into
account scale invariance and other symmetry elements arising as a result
of the fluctuations near the critical point (Ref. 3, Chap. 9, $\S$2).
The Landau theory is exact, giving an example of a complete symmetry theory,
only in high-dimensional spaces where the additional symmetry associated
with fluctuations does not arise. Another example is the conformal theory
of phase transitions for the two-dimensional case [4], which, proceeding
from the conformal invariance of the system at the critical point and the
finiteness of the number of strongly fluctuating quantities, fixes a
discrete set of possible values for critical exponents.

In the present paper we adopt the symmetry approach to the investigation
of the Anderson transition [5 -- 10], making a separation of the type (1)
not for the Hamiltonian $H$ but for an operator $\hat L$ which is the
quantum analog of the Boltzmann collision operator. The theory is based on
the following initial assumptions.

{\bf 1.} The Schr${\rm {\ddot o}}$dinger equation in a space of dimension
$d$
$$
[ \epsilon ({\bf {\hat p}}) + V ({\bf r})] \psi ({\bf r}) = E \psi ({\bf
r}) \eqno(2)
$$
describing the motion of non-interacting electrons with an arbitrary
spectrum $\epsilon ({\bf p})$ in a random potential $V ({\bf r})$ is
studied. As for the random potential, it is assumed only that the
averages with respect to its realizations can be calculated by the
diagrammatic technique. The existence of a diagrammatic technique for the
standard models of a random potential can be proved directly [1, 11, 12]. In
the general case,  the limits of applicability of the diagrammatic approach
are poorly investigated: some problems certainly arise for quasi-random
systems [13, 14].

The exact Green's function of Eq. (2) is expressed in terms of the
eigenfunctions $\psi_s ({\bf r})$ and eigenvalues $\epsilon_s$ $(s = 1, 2,
..., N)$,
$$
G_E^{R, A} ({\bf r}, {\bf r}^\prime) = \sum \limits_s \frac{\psi_s ({\bf
r}) \psi_s^* ({\bf r}^\prime)}{E - \epsilon_s \pm i \delta}.\eqno(3)
$$
The averaged Green's function $\langle G (r, r^\prime) \rangle$ is
determined by a diagrammatic series (Fig. 1a), and in accordance with
\begin{figure}
\centerline{\includegraphics[width=5.1 in]{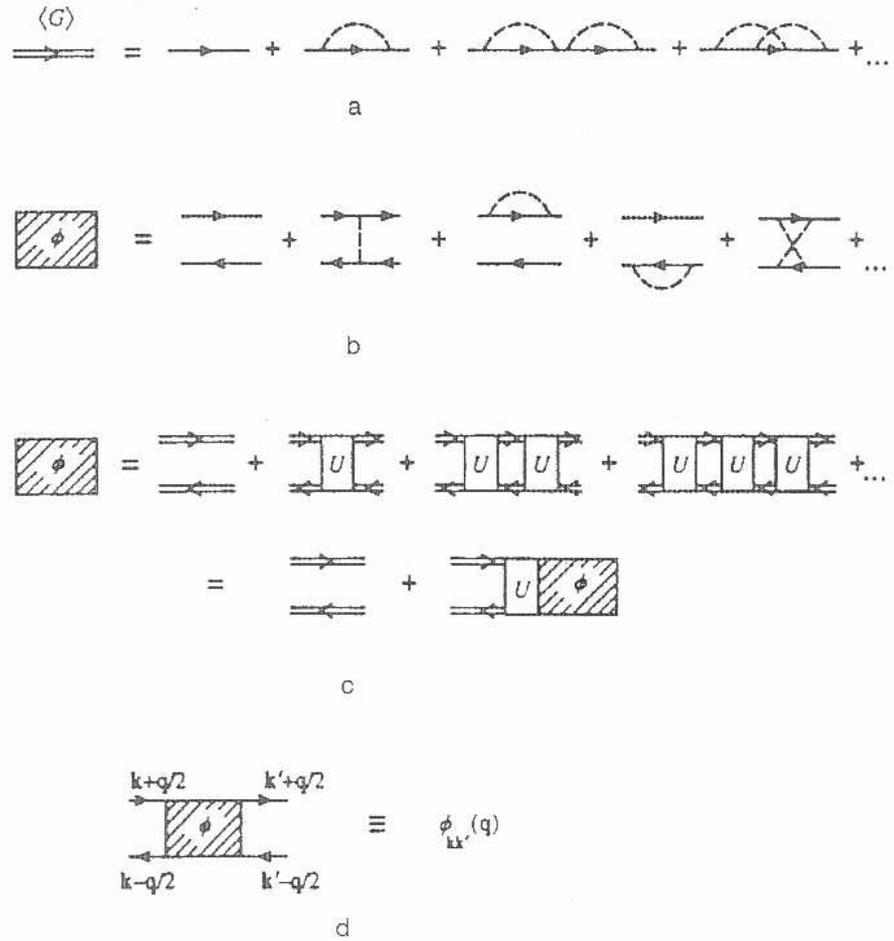}} \caption{
\,\, a, b ---  Diagrams for the average Green's
function (a) and the
quantity $\phi$ (b), which correspond to a Gaussian random
potential [1] or a Born approximation for randomly distributed
impurities [11] (in what follows, their specific form is not
used). c  ---  Graphical representation of the Bethe--Salpeter
equation; d  ---  explanation of the three-momentum notation. }
\label{fig1}
\end{figure}
current ideas [8, 15] it is assumed to be analytic at the point of the
Anderson transition: For $d \geq 4$ this was recently proved by the
present author [16, 17]. The quantity
$$
\phi ({\bf r}_1 {\bf r}_2, {\bf r}_3 {\bf r}_4) = \langle G_{E + \omega}^R
({\bf r}_1 {\bf r}_2) G_E^A ({\bf r}_3 {\bf r}_4) \rangle\eqno(4)
$$
which contains information about the kinetic properties, has a singularity
at the transition point. This quantity is determined by a series of
diagrams with four legs, constructed on $G^R$ and $G^A$ lines (Fig. 1b),
and its properties are similar to those of the two-paricle Green's
function in the theory of interacting particles [11]. It satisfies the
Bethe--Salpeter equation, containing an irreducible vertex $U$ (Fig. 1c).

{\bf 2.} The following symmetry elements are assumed:

(a) {\em Spatial uniformity in the mean.} This leads to a conservation law
for the external momenta in the diagrams. This makes it possible to
express $\langle G \rangle$ in terms of the self-energy $\Sigma$
$$
\langle G_E^{R,A} ({\bf k}) \rangle \equiv G_{\bf k}^{R,A} = \frac{1}{E -
\epsilon_{\bf k} - \Sigma_{\bf k}^{R,A}},\eqno(5)
$$
and to introduce for the function $\phi$ the three-momentum notation
$\phi_{{\bf kk}^\prime} ({\bf q})$ (Fig. 1d) and to write the
Bethe--Salpeter equation (Fig. 1c) in the form
$$
\phi_{{\bf kk}^\prime} ({\bf q}) = G_{{\bf k}+{\bf q}/2}^R G_{{\bf k}-{\bf
q}/2}^A \{ N \delta_{{\bf k-k}^\prime} +
 \frac{1}{N} \sum \limits_{{\bf k}_1} U_{{\bf kk}_1} ({\bf q}) \phi_{{\bf
k}_1 {\bf k}^\prime} ({\bf q}) \} .\eqno(6)
$$
Here and below the energy variable is equal to $E + \omega$ for the
functions $G^R$ and $E$ for the functions $G^A$.

(b) {\em Isotropy and inversional invariance in the mean}. With this
symmetry taken into account, the function $\phi_{{\bf kk}^\prime} ({\bf q})$
depends only on scalar products constructed from ${\bf k, k}^\prime$ and
${\bf q}$, whence, specifically,
$$ \phi_{{\bf k} {\bf k}^\prime} ({\bf q}) =
\phi_{- {\bf k}, - {\bf k}^\prime} (- {\bf q}).\eqno(7)
$$
Similarly, $G_{\bf k}^R$ and $G_{\bf k}^A$ depend on ${\bf k}^2$
and are even functions of $\bf k$.

(c) {\em Time-reversal invariance.} This property makes it possible to
choose real eigenfunctions $\psi_s ({\bf r})$ and to drop the
conjugation sign in Eq. (3). Then $G ({\bf r}, {\bf r}^\prime) = G ({\bf
r}^\prime, {\bf r})$ and interchanging ${\bf r}_1, {\bf r}_2$ and ${\bf
r}_3, {\bf r}_4$ in Eq. (4) gives in the momentum
representation
$$
\phi_{{\bf k} {\bf k}^\prime} ({\bf q}) = \phi_{- {\bf k}^\prime, - {\bf
k}} (- {\bf q})\eqno(8)
$$
$$
\phi_{{\bf k} {\bf k}^\prime} ({\bf q}) = \phi_{({\bf k} - {\bf k}^\prime +
{\bf q})/2, ({\bf k}^\prime - {\bf k} + {\bf q})/2} ({\bf k} + {\bf
k}^\prime) \eqno(9)
$$
Comparing Eqs. (7) and (8), we obtain
$$
\phi_{{\bf k} {\bf k}^\prime} ({\bf q}) = \phi_{{\bf k}^\prime {\bf k}}
({\bf q}).\eqno(10)
$$
Solving Eq. (6) formally for the function $U_{{\bf kk}^\prime}
({\bf q})$ and using Eqs. (7) and (10), it is easy prove similar
properties for this function:
$$
U_{{\bf k} {\bf k}^\prime} ({\bf q}) = U_{- {\bf k}, - {\bf k}^\prime} (-
{\bf q}), \, \, \, \, \, \, U_{{\bf k} {\bf k}^\prime} ({\bf q}) = U_{{\bf
k}^\prime {\bf k}} ({\bf q}).\eqno(11)
$$

{\bf 3.} It is conventionally assumed that the Anderson transition occurs
from a phase with exponential localization of the wave functions into a
phase with a finite diffusion coefficient. The existence of exponential
localization in the limit $E \rightarrow - \infty$ and finite diffusion
for large positive values of $E$ (for $d > 2$ and an unbounded spectrum
$\epsilon ({\bf p}), \, \, 0 \leq \epsilon ({\bf p}) \leq \infty$) has
been firmly established, as a result of many investigations, for Eq. (2).
The proof of the existence of a mobility edge is based mainly on
Mott's argument [7]: The existence of states with different degree of
localization and the same energy is impossible because of instability with
respect to an infinitesimal perturbation of general type. Mott's argument
does not forbid, however, the existence of intermediate states --- with
power-law localization, hybrid states, and so on --- and correspondingly
different types of "Anderson transitions" (for example, in the quasi-random
systems [13, 14] the transition occurs from exponential localization to a
ballistic regime). In the present paper the first instability, arising
with a motion from deep in an exponentially localized phase, is
investigated and it is shown that it does indeed correspond to a
transition into a phase with finite diffusion.

{\bf 4.} The general notions of the modern theory of critical phenomena
[1] (such as parameter space, critical surface, relevant and
irrelevant parameters) are used.

{\bf 5.} The theory is based on the physical idea that the localization
phenomenon is associated with a diffusion pole in the irreducible four-leg
vertex
$$
U_{{\bf k} {\bf k}^\prime} ({\bf q}) = U_{{\bf k} {\bf k}^\prime}^{reg}
({\bf q}) + U_{{\bf k} {\bf k}^\prime}^{sing} ({\bf q}) = U_{{\bf k} {\bf
k}^\prime}^{reg} ({\bf q}) +
 \frac{F ({\bf k}, {\bf k}^\prime, {\bf
q})}{-i \omega + D (\omega, {\bf k} + {\bf k}^\prime)({\bf k} + {\bf
k}^\prime)^2}\eqno(12)
$$
proposed by Vollhardt and W${\rm {\ddot o}}$lfle in the so-called
"self-consistent theory of localization" (see Ref. 18, and also Refs. 10
and 19).

This idea agrees with the theory of weak localization [20 -- 22],
according to which the diffusion pole in $U_{{\bf kk}^\prime} ({\bf q})$
determines the main quantum corrections to the conductivity which in turn
determine the scaling behavior in a space with dimension $d = 2 +
\epsilon$. The diffusion pole in $U_{{\bf kk}^\prime} ({\bf q})$ with the
classical diffusion coefficient $D_{cl}$ arises as a result of summation
of fan-shaped diagrams [20]; Vollhardt and W${\rm {\ddot o}}$lfle
conjectured that when {\it all} diagrams are taken into account, $D_{cl}$ is
replaced by the exact diffusion coefficient $D (\omega, {\bf q})$.
They approximated $U_{{\bf kk}^\prime} ({\bf q})$ by Eq.~12 with $U_{{\bf
kk}^\prime}^{reg} ({\bf q}) = \mbox{const}, \, \, F ({\bf k}, {\bf k}^\prime,
{\bf q}) = \mbox{const}$ and solved approximately the Bethe--Salpeter
equation (6), which, using the Ward identity [18]
$$
\Delta \Sigma_{\bf k} ({\bf q}) = \frac{1}{N} \sum \limits_{{\bf k}_1}
U_{{\bf kk}_1} ({\bf q}) \Delta G_{{\bf k}_1} ({\bf q}),\eqno(13)
$$
$$
\Delta G_{\bf k} ({\bf q}) \equiv G_{{\bf k} + {\bf q}/2}^R - G_{{\bf k} -
{\bf q}/2}^A\,,  \qquad
\Delta \Sigma_{\bf k} ({\bf q}) \equiv \Sigma_{{\bf k} + {\bf q}/2}^R -
\Sigma_{{\bf k} - {\bf q}/2}^A\eqno(14)
$$
was rewritten in the form
$$
[- \omega + (\epsilon_{{\bf k} + {\bf q}/2} - \epsilon_{{\bf k} - {\bf
q}/2})] \phi_{{\bf k} {\bf k}^\prime} ({\bf q}) +
$$
$$
+\frac{1}{N} \sum
\limits_{{\bf k}_1} U_{{\bf k} {\bf k}^\prime} ({\bf q})
\, [\Delta G_{{\bf k}_1} ({\bf q}) \phi_{{\bf k} {\bf k}^\prime} ({\bf
q}) - \Delta G_{\bf k} ({\bf q}) \phi_{{\bf k}_1 {\bf k}^\prime} ({\bf
q})] =
 \Delta G_{\bf k} ({\bf q}) N \delta_{{\bf k} - {\bf k}^\prime}.\eqno(15)
$$
There exists a simple estimate which gives the same results. We note that
the second term in the left-hand side of Eq. (15) is reminiscent of a
Boltzmann collision integral and indeed transfer into it in the limit of weak
disorder (Sec. 3). It is significant that in the quantum region the quantity
$U_{{\bf kk}^\prime} ({\bf q})$ plays the role of a "transition probability".
Using analogue of $\tau$-approximation, $D \propto l \propto \langle U
\rangle ^{-1}$ ($l$ is the mean free path, $\langle ...  \rangle$ denotes
averaging over the momenta), and taking into account Eq.  (12), we obtain the
self-consistency equation of the Vollhardt--W${\rm {\ddot o}}$lfle theory
$$
D \sim {\rm const} \left ( U_0 + F_0 \int \frac{d^dq}{-i \omega +
D(\omega, q)q^2} \right )^{-1}.\eqno(16)
$$
This estimate is no less accurate and demonstrates more clearly the crux
of the matter than the approximate solution given in Ref. 18 for Eq. (15).
As the degree of disorder increases, the "transition probability"
increases anomalously as a result of a decrease of the diffusion
coefficient, making it possible for this coefficient to vanish. Neglecting
the spatial dispersion $D (\omega, {\bf q})$, Eq. (16) makes it possible
to determine the critical exponents for the conductivity $\sigma$ and the
localization length $\xi$
$$
\sigma \sim \tau^s, \qquad \, \, \, \xi \sim \tau^{- \nu}\eqno(17)
$$
($\tau$ is the distance to the transition); setting $D = \mbox{const}
(\omega) \sim \sigma$ in the metal phase and $D \sim (-i \omega) \xi^2$ in
the localized phase, we obtain
$$
s = 1\,\,, \quad d > 2\,\,;
 \qquad \qquad  \nu = \left \{ \begin{array}{cc}
{\displaystyle \frac{1}{d-2}}\quad , & 2 < d < 4 \\ {   } & {  } \\
{\displaystyle \frac{1}{2}} \quad, & d > 4
\end{array} \right . \quad.
\eqno(18)
$$

The drawbacks of the self-consistent theory [18]
can already be seen from the exposition given above:

(a) The method used to solve the Bethe--Salpeter equation is rough;

(b) The spatial dispersion $D (\omega, { q})$ is ignored, while
it  can change
substantially the estimate of the integral in Eq. (16);

(c) An approximation is used for $U_{{\bf kk}^\prime} ({\bf q})$ that leads
to a singularity $\sim 1/ \omega$ on the right-hand side of Ward's
identity (13) in the localized phase, and this is incompatible with
regularity of $\Sigma$ at the transition point.

One of the most interesting questions in the theory of localization is
connected with the drawback (b). It follows from the Berezinskii--Gor'kov
criterion [24] that in the localized phase $D (0, { q}) \equiv 0$ (Sec.
4). The question arises, how the spatial dispersion of $D$ changes
near the Anderson transition. Vollhardt and W${\rm {\ddot o}}$lfle
had in mind  that $D (0, { q})$ vanishes at the transition
point simultaneously for all values of $q$; in the clear form, such
hypothesis was stated by Efetov [21]. The vanishing of the whole function
cannot occur accidently and it must be supported by a deep symmetry. Does
this symmetry exist? What is its nature?

Another fundamental question of the
theory is touched upon in the drawback (c). If the diffusion pole in $U_{{\bf
kk}^\prime} ({\bf q})$ exists, then why is there no $ 1/ \omega$
singularity on the right-hand side of Ward's identity (13)? The condition for
this pole to be cancelled imposes stringent requirements on the approximation
employed, while the satisfaction of Ward's identity has actually never been
checked in any of the existing theories [10].

A theory free of the drawbacks (a--c) and answering the questions raised is
expounded below. The first half of this paper follows the scheme of the
Vollhardt--W${\rm {\ddot o}}$lfle theory and contains a proof of the
relations obtained in Ref. 18 by a chain of hypotheses or doubtful
approximations. In Secs. 2 and 3 the diffusion poles of $\phi_{{\bf
kk}^\prime} ({\bf q})$ are separated out and the result (12) is proved. In
Sec. 4 the general properties of the diffusion coefficient and its
relation with the localization of the wave functions are determined. The
content of Secs. 5--7 replaces the rough solution of the Bethe--Salpeter
equation [18]: In Sec. 5 a hierarchical structure of the spectrum is
obtained for the quantum collision operator $\hat L$; a separation of the
type (1), convenient for a symmetry analysis, is established;  a
condition on the transition point is found and a self-consistency
equation, replacing Eq. (16), is derived. The diffusion coefficient $D
(\omega, {q})$ is sought with no assumptions on the character of the
spatial dispersion (Secs. 6 and 7), but only the solution with a weak
$q$ dependence is found to be internally consistent. Such weak
dependence does not affect the estimate of the
integral in Eq. (16) and leads to the result (18) for the critical
exponents. In summary, all basic results of Ref. 18 are found
to be correct, which is surprising for such a rough theory.

The theory expounded starts from the obvious symmetry of the system and
the additional symmetry of the critical point is determined in the course
of the analysis. The inevitable question is: Are the hidden symmetry
elements completely determined? There is a serious argument,  that
the determination is complete (Sec. 8): with respect to the character
of the change in symmetry, the Anderson transition is found to be similar to
the Curie point for an isotropic $n$-component ferromagnet in the limit $n
\rightarrow \infty$. This model of a ferromagnet is the basis of the $1/n$
expansion [1], its critical exponents are known exactly, and they are in
exact agreement with Eq. (18). The isotropy of the equivalent ferromagnet
is the symmetry that makes $D (0, {q})$ vanish simultaneously for all
$q$. The approximate (to an accuracy $\sim \omega$) orthogonality of
the singular part of $U_{{\bf kk}^\prime} ({\bf q})$ and $\Delta G_{\bf k}
({\bf q})$ results in cancellation of singularity in the
right-hand side of the Ward identity [Sec. 5.3, Eq. (13)].

Another method for checking the completeness of the symmetry found is to
compare with the results of model investigations. The hypothesis that the
exponents (18) are exact was actually stated in Ref. 25 on the basis of an
analysis of all known results:

(a) For $d = 2 + \epsilon$ Wegner's relation $s = (d - 2) \nu$, following
from the existence of one-parameter scaling [21], is valid and the
$\epsilon$ expansion for the exponent $\nu$ has the form [26]
$$
\nu = \frac{1}{\epsilon} + 0 \cdot \epsilon^0 + 0 \cdot \epsilon^1 + O
(\epsilon^2),\eqno(19)
$$
which agrees with Eq. (18), if the coefficients of the higher order powers
of $\epsilon$ are also zero.

(b) The result (18) separates  the dimensions of the space $d_{c1} = 2$ and
$d_{c2} = 4$, which on the basis of independent analysis are
considered to be the lower [21] and the upper (see the discussion and
references in Ref. 16) critical dimensions.

(c) The entire experience of the theory of phase transitions shows that
for $d > d_{c2}$ the critical exponents do not depend on $d$, which is the
case in Eq. (18).

(d) The exponents (18) agree with the results for $d = \infty : \, \, \nu
= 1/2$ (Refs. 27 and 28) and $s = 1$ (Ref. 29); the disagreement with the
result $s = \infty$, obtained in Ref. 28, is discussed in Sec. 9.

The value $\nu = 1$ of the exponent for $d = 3$ agrees satisfactorily with
the results of numerical calculations ($\nu = 1.2 \pm 0.3$ (Ref. 30) and
$\nu = 1.5 \pm 0.2$ (Ref. 31) and the qualitatively behavior of $\nu$ as a
function of $d$ agrees with the estimates from hierarchical models [14].
In Wegner's work [32] a finite contribution $\sim \epsilon^2$ is obtained
in Eq. (19). This makes the agreement with the numerical calculations of
Refs. 30 and 31 much worse. However, this result was derived for the
zero-component $\sigma$-model, whose correspondence with the initial
disordered system is controversial (Wegner himself [32] does not reject
this), and it is apparently correct only in the lowest orders in $\epsilon$
(Sec.  9).

A qualitative result of this work, which can be checked experimentally, is
the assertion that there is no spatial dispersion of $D (\omega, {q})$
on the scale $\xi^{-1}$ in the limit $\omega \to 0$ (compare with Refs. 33,
34, and 35).  The absence of
significant spatial dispersion does not contradict the strong dependence of
the diffusion coefficient $D_L$ of a finite system on its size
$L$ [21]. This dependence is connected with the temporal dispersion and is
determined in terms of the known function $D (\omega, q)$ by
the relation $D_L \sim D(D_L /L^2, 0)$ [18].
\\

\begin{center}
{\bf 2. RELATIONS BETWEEN THE QUANTITIES IN THE PRESENCE OF SPATIAL AND
TEMPORAL DISPERSION} \\
\end{center}

In this section the existence of a diffusion pole in the quantity
$$
\phi ({\bf q}) = \frac{1}{N^2} \sum \limits_{{\bf kk}^\prime} \phi_{{\bf
kk}^\prime} ({\bf q}),\eqno(20)
$$
which is the Fourier transform of the quantity (4) for coinciding
arguments ${\bf r}_1 = {\bf r}_4, \, \, {\bf r}_2 = {\bf r}_3$, will be
proved. In contrast to Ref. 18 and other works, it will not be assumed
that $q$ is small. In view of the great confusion in the literature, we
shall give a complete summary of the formulas which are relevant here.

We shall consider the response of a system to an electric field
${\bf E} ({\bf r}, t) \sim e^{i {\bf q} \cdot {\bf r} - i \omega
t}$. The frequency $\omega$ is assumed to be finite only in order
to remove the uncertainties which appear; the limit $\omega
\rightarrow 0$ is taken in the final results. Neglecting magnetic
effects, the field ${\bf E}$ is a purely potential field. This
makes it possible to confine attention to the longitudinal
components of the susceptibilities (Ref. 36, $\S$103). In the
presence of spatial dispersion two definitions of the conductivity
are possible:
$$
\begin{array}{c}
{\bf j} (\omega, {\bf q}) = {\tilde \sigma} (\omega, {\bf q})
{\bf E} (\omega, {\bf q})           \\
{\bf j}_e (\omega, {\bf q}) = \sigma (\omega, {\bf q}) {\bf E}
(\omega, {\bf q})
\end{array}\,\quad\qquad
{\bf j} = {\bf j}_e + {\bf j}_{diff} , \eqno(21)
$$
which relate ${\bf E}$ with the total current ${\bf j}$ or its
electric component ${\bf j}_e$; the diffusion component of the
current ${\bf j}_{diff} (\omega, {\bf q}) = -i {\bf q} D (\omega,
{\bf q}) \rho (\omega, {\bf q})$ is due to the deviation of the
electron density $\rho$ from the equilibrium density. This
deviation is determined by the polarizability $\alpha$ ($\varphi$
is a scalar potential):
$$
\rho (\omega, {\bf q}) = \alpha (\omega, {\bf q}) \varphi (\omega, {\bf
q}) .\eqno(22)
$$
The conductivity $\tilde \sigma$ appears in Kubo's formulas (see below),
which determine the total response of the system to the field ${\bf E}$.
The conductivity $\sigma$ is related with the diffusion coefficient $D$ by
the Einstein relation
$$
\sigma (\omega, {\bf q}) = e^2 N(\epsilon_F) D (\omega, {\bf q})\eqno(23)
$$
since the change in the scalar potential $\varphi$ and the chemical
potential $\mu$ with $\mu ({\bf r}, t) + e \varphi ({\bf r}, t) =
\mbox{const}$ does not destroy the thermodynamic equilibrium ($N
(\epsilon_F)$ is the density of states at the Fermi level). A relation
between $\sigma, \, \, {\tilde \sigma}$, and $\alpha$ follows from the
continuity equation:
$$
-i \omega {\tilde \sigma} (\omega, {\bf q}) = [-i \omega + D (\omega, {\bf
q}) q^2] \sigma (\omega, {\bf q}),
$$
$$
\omega \alpha (\omega, {\bf q}) = -iq^2 {\tilde \sigma} (\omega, {\bf
q}),\eqno(24)
$$
so that the difference between $\sigma$ and $\tilde \sigma$ is important
only for ${\bf q} \neq 0$. Using the relation (23), we obtain for the
polarizability $\alpha$ and the permittivity $\epsilon$
$$
\alpha (\omega, {\bf q}) = - \frac{e^2 N(\epsilon_F) D (\omega, {\bf q})
q^2}{-i \omega + D (\omega, {\bf q}) q^2},
$$
$$
\epsilon (\omega, {\bf q}) = 1 - \frac{4 \pi}{q^2} \alpha (\omega, {\bf
q}) = 1 - \frac{4 \pi {\tilde \sigma} (\omega, {\bf q})}{i \omega}
.\eqno(25)
$$
It is clear from Eqs. (23 -- 25) that if the diffusion coefficient $D
(\omega, {\bf q})$ is given, then all quantities introduced above can be
determined.

The quantities $\tilde \sigma$ and $\alpha$ are given by Kubo's formulas
(Ref. 2, $\S$126, and Ref. 37, $\S$75)
$$
{\tilde \sigma} (\omega, {\bf q}) = \frac{1}{\omega} \int \limits_0^\infty
dt e^{i \omega t} \int d{\bf r} e^{-i{\bf q} \cdot {\bf r}} \langle
{\hat j} ({\bf r}, t) {\hat j} (0, 0) -
 {\hat j} (0, 0) {\hat j} ({\bf r} t) \rangle ,\eqno(26)
$$
$$
{\tilde \sigma} (\omega, {\bf q}) = \frac{1}{q} \int \limits_0^\infty
dt e^{i \omega t} \int d{\bf r} e^{-i{\bf q} \cdot {\bf r}} \langle
{\hat j} ({\bf r}, t) {\hat \rho} (0, 0) -
 {\hat \rho} (0, 0) {\hat j} ({\bf r}, t) \rangle ,\eqno(27)
$$
$$
\alpha (\omega, {\bf q}) = -i \int \limits_0^\infty dt e^{i \omega t} \int
d{\bf r} e^{-i{\bf q} \cdot {\bf r}} \langle {\hat \rho} ({\bf r}, t)
{\hat \rho} (0, 0) -
 {\hat \rho} (0, 0) {\hat \rho} ({\bf r}, t) \rangle\eqno(28)
$$
which determine, respectively, the response of the current to a vector
potential, the response of a current to the scalar potential, and the
response of the density to a scalar potential. The equivalence of Eqs.
(26) and (27) and the relation (24) between $\tilde \sigma$ and $\alpha$
follow from the continuity equation for the density operator $\hat \rho$
and the longitudinal component of the current operator $\hat j$ and the
asymptotic expressions for $\tilde \sigma$ and $\alpha$ in the limit
$\omega \rightarrow \infty$ (Ref. 36, $\S$78).

We note that according to the precise meaning of Kubo's formula (see the
detail discussion in Ref. 38), the response of the system to the field
$\bf D$ produced by external charges must be calculated. In this approach
the Coulomb interaction between the electrons must be necessarily included
in the Hamiltonian to avoid contradictions in the Maxwell's equations;
Kubo's formulas have a form that is somewhat different from Refs. 26--28
(Ref. 37, p. 413), and the correlation functions appearing in them must be
calculated taking into account the Coulomb interaction. A different
approach [38] is more convenient: The interaction between the electrons is
divided into a short-range and slowly-varying long-range parts; the first
part is included explicity in the Hamiltonian and the second part is taken
into account as a self-consistent field, leading to screening of the field
$\bf D$; for this reason, the response to a real physical field $\bf E$ is
studied and the correlation functions appearing in Eqs. (26--28) are
calculated only taking into account the short-range part of the
interaction. The latter part can be taken into account in the spirit of
the Fermi-liquid theory. We shall neglect it completely, since in its
classical formulation the Anderson transition problem is a problem of
noninteracting electrons. We emphasize,  that the word
"noninteracting" must be understood precisely in the sense indicated above,
since otherwise the concept of conductivity cannot be introduced in a
consistent manner.

The correlation function in Eq. (28) for noninteracting electrons in a
random potential is calculated similarly to the correlation function for a
Fermi gas (Ref. 2, $\S$117) using, instead of the plane-wave
representation, a representation in terms of the eigenfunctions $\psi_s
(r)$ of Eq. (2)\footnote{For definiteness, all quantities refer to the
same spin projection. For purely potential scattering the spin subsystems
are independent and the number of spin components can be easily taken into
account in the final results.}:
$$
\alpha (\omega, {\bf q}) = e^2 \int \limits_{- \infty}^\infty d \epsilon
\int \limits_{- \infty}^\infty d \omega^\prime \frac{f_0 (\epsilon) -
f_0 (\epsilon + \omega^\prime)}{\omega - \omega^\prime + i \delta} N
(\epsilon )
\, \langle \rho_\epsilon \rho_{\epsilon + \omega^\prime} \rangle_{\bf
q}.\eqno(29)
$$
Here $f_0 (\epsilon)$ is the Fermi function, and $\langle \rho_E \rho_{E +
\omega} \rangle_{\bf q}$ is the Fourier transform of the
Berezinskii--Gor'kov spectral density [24]
$$
\langle \rho_E ({\bf r}) \rho_{E + \omega} ({\bf r}^\prime) \rangle =
\frac{1}{N(E)} \langle \sum \limits_{ss^\prime} \psi_s^* ({\bf r})
\psi_{s^\prime} ({\bf r}) \psi_{s^\prime}^* ({\bf r}^\prime)
\, \psi_s ({\bf r}^\prime) \delta (E - \epsilon_s) \delta (E -
\epsilon_{s^\prime} + \omega ) \rangle .\eqno(30)
$$
For small $\omega$ and zero temperature, taking the imaginary part of Eq.
(29), we obtain the inversion of (29)
$$
\langle \rho_{\epsilon_F} \rho_{\epsilon_F + \omega^\prime} \rangle_{\bf
q} = - \frac{\mbox{Im} \alpha_{\epsilon_F} (\omega, {\bf q})}{\pi e^2
\omega N(\epsilon_F)}\eqno(31)
$$
(in the absence of interaction $\psi_s ({\bf r})$ and $\epsilon_s$ do not
depend on $\epsilon_F$ and $\epsilon_F$ can be replaced by $E$). The
standard diffusion form for $\langle \rho_E \rho_{E + \omega}
\rangle_{\bf q}$ (Refs. 10 and 24) is obtained by substituting the
expression (25) into Eq. (31), assuming $D (\omega, {\bf q})$ is real,
which in general is not the case.

The following expression can be easily obtained for the function $\phi
({\bf q})$ ($E$ and $\omega$ are parameters appearing in Eq. (4)):
$$
\phi ({\bf q}) = \int \limits_{- \infty}^\infty d \epsilon \int \limits_{-
\infty}^\infty d \omega^\prime \frac{N (\epsilon) \langle
\rho_\epsilon \rho_{\epsilon + \omega^\prime} \rangle_{\bf q}}{(E + \omega
- \epsilon + i \delta)(E - \omega^\prime - \epsilon - i \delta)}.\eqno(32)
$$
The polarizability $\alpha (\omega, {\bf q})$ is a generalized
susceptibility (Ref. 2, $\S$123; Ref.36, $\S$103) and the oddness of
$\mbox{Im} \alpha (\omega, {\bf q})$ as a function of the frequency makes
it possible to write
$$
\mbox{Im} \alpha (\omega, {\bf q}) = \frac{{\tilde \alpha} (\omega, {\bf
q}) - {\tilde \alpha} (- \omega, {\bf q})}{2i}\,, \qquad
{\tilde \alpha} (\omega, {\bf q}) = \alpha (\omega, {\bf q}) - \alpha (0,
{\bf q}).\eqno(33)
$$
Substituting the expressions (31) and (33) into Eq. (32), we obtain
integrals with ${\tilde \alpha} (\omega^\prime, {\bf q})$ and ${\tilde
\alpha} (- \omega^\prime, {\bf q})$ that converge separately. Making the
substitution $\omega^\prime \rightarrow - \omega^\prime$ in the second of
the integrals and shifting upwards the contour integration over
$\omega^\prime$, and taking into account the fact that $\alpha (\omega,
{\bf q})$ is analytic in the upper half-plane, we obtain
$$
\phi ({\bf q}) = \frac{1}{e^2} \int \limits_{- \infty}^\infty d
\omega^{\prime \prime} \frac{{\tilde \alpha}_{E +
\omega^{\prime \prime}} (\omega^{\prime \prime}, {\bf q})}{(\omega^{\prime
\prime} - \omega - i \delta)(\omega^{\prime \prime} + i \delta)} \simeq
 \frac{2 \pi i}{e^2 \omega} {\tilde \alpha}_E (\omega, {\bf
q}),\eqno(34)
$$
where $\omega^{\prime \prime} = \epsilon - E$. The second equality follows
by neglecting $\omega^{\prime \prime}$ in the argument $E + \omega^{\prime
\prime}$, as a function of which appreciable changes in $\tilde \alpha$
occur on the atomic scale and are not important in the region
$\omega^{\prime \prime} \sim \omega$, which makes the main contribution to
the integral. Substituting $\alpha (\omega, {\bf q})$ in the form (25), we
obtain
$$
\phi ({\bf q}) = \frac{2 \pi N (E)}{-i \omega + D(\omega, {\bf q}) q^2} +
\phi_{reg} ({\bf q}),\eqno(35)
$$
where the contribution $\phi_{reg} ({\bf q})$ originates from the region
of large values of $\omega^{\prime \prime}$ in Eq. (34) and is regular in
the limit $\omega, {\bf q} \rightarrow 0$. In the localized phase, when $D
(\omega, {\bf q}) \sim (- i \omega)$ (Sec. 4), to obtain the expression
(35)  we should add a small real frequency-independent term in the
denominator, in order for all expressions to be meaningful; it is
essential in separating $\alpha (0, {\bf q})$ from
$\alpha (\omega, {\bf q})$. In conclusion, the quantity $\phi ({\bf q})$ has a
diffusion pole which contains the observable diffusion coefficient. \\

\begin{center}
{\bf 3. SEPARATION OF DIFFUSION POLES FROM THE BETHE--SALPETER EQUATION}
\\
\end{center}

We introduce the operator $\hat L$, which is the symmetrized version of
the operator on the left-hand side of Eq. (15), which arises as a result
of the replacement $\phi_{{\bf kk}^\prime} ({\bf q}) \rightarrow
\phi_{{\bf kk}^\prime} ({\bf q}) \sqrt {\Delta G_{\bf k} ({\bf q})}$ and
division of Eq. (15) by $\sqrt {\Delta G_{\bf k} ({\bf q})}$:
$$
{\hat L} ({\bf q}) = {\hat L}_0 ({\bf q}) + {\hat M} ({\bf q})\eqno(36)
$$
$$
{\hat L}_0 \psi_{\bf k} \equiv \frac{1}{N} \sum \limits_{{\bf k}_1}
U_{{\bf kk}_1} ({\bf q}) [ \Delta G_{{\bf k}_1} ({\bf q}) \psi_{\bf k} -
 \sqrt{\Delta G_{\bf k} ({\bf q}) \Delta G_{{\bf k}_1} ({\bf q})}
\psi_{{\bf k}_1}],
$$
$$
{\hat M} \psi_{\bf k} \equiv (\epsilon_{{\bf k} + {\bf q}/2} -
\epsilon_{{\bf k} - {\bf q}/2}) \psi_{\bf k}.
$$
The operator $\hat L$ acts in the complex space and, by virtue of Eq.
(11), it is symmetrical with respect to the scalar product
$$
(\phi, \psi) = \frac{1}{N} \sum \limits_{\bf k} \phi_{\bf k} \psi_{\bf k}
.\eqno(37)
$$
Its eigenfunctions $e_{\bf k}^{(s)} ({\bf q})$ form a complete orthonormal
basis, and the eigenvalues $\lambda_s ({\bf q})$ are, generally speaking,
complex. In terms of $\lambda_s$ and $e^{(s)}$ the formal solution of the
Bethe--Salpeter equation (15) has the form
$$
\phi_{{\bf kk}^\prime} ({\bf q}) = \sum \limits_s \frac{f_{\bf k}^{(s)}
({\bf q}) f_{{\bf k}^\prime}^{(s)} ({\bf q})}{- \omega + \lambda_s ({\bf
q})},
$$
$$
f_{\bf k}^{(s)} ({\bf q}) = \sqrt{\Delta G_{\bf k} ({\bf q})} e_{\bf
k}^{(s)} ({\bf q}).\eqno(38)
$$
At least, one eigenvalue --- for definiteness $\lambda_0 ({\bf q})$ ---
behaves as $\lambda_0 ({\bf q}) \sim q^2$ for small $\bf q$. Indeed, the
operator ${\hat L}_0$ has a zero mode $\psi_{\bf k} ({\bf q}) = \sqrt {\Delta
G_{\bf k} ({\bf q})}$ and, treating the operator ${\hat M} \sim q$ as a
perturbation, we can construct the iterative series

$$
e_{\bf k}^{(0)} ({\bf q}) = \mbox{const} [\psi_{\bf k}^{(0)} ({\bf q}) +
\psi_{\bf k}^{(1)} ({\bf q}) + ...],
$$
$$
\psi_{\bf k}^{(0)} ({\bf q}) = \sqrt{\Delta G_{\bf k} ({\bf q})},
$$
$$
\lambda_0 ({\bf q}) = \lambda_0^{(1)} ({\bf q}) + \lambda_0^{(2)} ({\bf
q}) + ..., \qquad  \lambda_0^{(n)} ({\bf q})\,,\quad \psi^{(n)} \sim
q^n,\eqno(39)
$$
$$
(\psi^{(0)}, \psi^{(n)}) = 0\,,\qquad n \neq 0
$$
in the Brillouin--Wigner form [39]. The eigenvalues $\lambda_s ({\bf q})$
are even with respect to $\bf q$ (see Appendix) and the correction
$\lambda_0^{(1)}$ is equal to zero, as one can easily verify directly.
To second order in $q$ we have
$$
\lambda_0 ({\bf q}) = \frac{(\psi^{(0)}, {\hat M} \psi^{(0)}) +
(\psi^{(0)}, {\hat M} \psi^{(1)})}{(\psi^{(0)}, \psi^{(0)})}\eqno(40)
$$
where $\psi^{(1)}$ satisfies the equation
$$
- {\hat P}_\bot {\hat M} \psi^{(0)} = {\hat L}_0 \psi^{(1)}\eqno(41)
$$
and ${\hat P}_\bot$ is a projection operator onto the space orthogonal to
$\psi^{(0)}$ (Ref. 39). Making the substitution
$$
\psi_{\bf k}^{(1)} ({\bf q}) = - i \sqrt {\Delta G_{\bf k} ({\bf q})} {\bf
q} \cdot {\bf l}_{\bf k}\eqno(42)
$$
and noting that
$$
\frac{1}{N} \sum \limits_{\bf k} \Delta G_{\bf k} ({\bf q}) = \frac{1}{N}
\sum \limits_{\bf k} 2i \mbox{Im} G_{\bf k}^R = - 2 \pi i N(E)\eqno(43)
$$
we rewrite (40) in the form (${\bf v}_{\bf k}$ is the velocity of
electrons with momentum $\bf k$)
$$
\lambda_0 ({\bf q}) = \frac{i}{2 \pi N(E)} \left [ \frac{1}{N} \sum
\limits_{\bf k} ({\bf q} \cdot {\bf v}_{\bf k}) ({\bf q} \cdot {\bf
l}_{\bf k}) (-i) \Delta G_{\bf k} ({\bf q}) +
 \frac{1}{N} \sum \limits_{\bf k} ({\bf q} \cdot {\bf v}_{\bf k})
\Delta G_{\bf k} ({\bf q}) \right ] .\eqno(44)
$$
For an isotropic spectrum $\epsilon ({\bf k}) = k^2/2m$ the expressions
(44) and (41) assume, to lowest order in $\bf q$, the form
$$
\lambda_0 ({\bf q}) = -iD(0,0) q^2, \, \, \, \, D (0, 0) = \sigma (0, 0)
e^{-2} N^{-1} (E),\eqno(45)
$$
$$
\sigma (0, 0) = \frac{e^2}{2 \pi d} \frac{1}{N} \sum \limits_{\bf k} ({\bf
v}_{\bf k} \cdot {\bf l}_{\bf k}) i \Delta G_{\bf k} (0) +
 \frac{e^2}{2 \pi m} \frac{1}{N} \sum \limits_{\bf k} \mbox{Re} G_{\bf
k}^R,\eqno(46)
$$
$$
{\bf v}_{\bf k} = \frac{1}{N} \sum \limits_{{\bf k}^\prime} iU_{{\bf
kk}^\prime} (0) \Delta G_{{\bf k}^\prime} (0) ({\bf l}_{\bf k} - {\bf
l}_{{\bf k}^\prime}).\eqno(47)
$$
In the limit of weak disorder, when
$$
\Delta G_{\bf k} (0) = G_{\bf k}^R - G_{\bf k}^A = 2i \mbox{Im}
\frac{1}{E - \epsilon_{\bf k} + i \gamma} \approx
 - 2 \pi i \delta (E - \epsilon_{\bf k})\eqno(48)
$$
we obtain from Eqs. (46) and (47)
$$
\sigma (0, 0) = \frac{e^2}{d} \, \frac{1}{N} \sum \limits_{\bf k} {\bf
v}_{\bf k} \cdot {\bf l}_{\bf k} \delta (E - \epsilon_{\bf k}),\eqno(49)
$$
$$
{\bf v}_{\bf k} = \frac{1}{N} \sum \limits_{{\bf k}^\prime} 2 \pi U_{{\bf
kk}^\prime} (0) ({\bf l}_{\bf k} - {\bf l}_{{\bf k}^\prime}) \delta (E -
\epsilon_{{\bf k}^\prime}),\eqno(50)
$$
i.e., $\sigma (0, 0)$ is the classical conductivity, $D (0, 0)$ is the
classical diffusion coefficient, and ${\bf l}_{\bf k}$ is the vector mean
free path length, determined by the standard classical equation (50) for
scattering by impurities [40]. The results (46) and (47) extend the
concept of a kinetic equation and a mean free path into the quantum
region. The differences from the classical equations reduce to the
following:

(a) The $\delta$-function expressing the energy conservation law is
smeared;

(b) the transition probability is replaced with $2 \pi U_{{\bf
kk}^\prime} (0)$;

(c) $\sigma$ acquires a quantum correction [last term in Eq. (46)] of the
order of the Mott minimum conductivity [7].

It is obvious from Eq. (45) that the diffusion pole is related with the
zeroth term in the sum in Eq. (38). To compare with Eq. (35), we sum the
expression (38) over $\bf k$ and ${\bf k}^\prime$:
$$
\phi ({\bf q}) = \frac{A_0 ({\bf q})^2}{- \omega + \lambda_0 ({\bf q})} +
\sum \limits_{s \neq 0} \frac{A_s ({\bf q})^2}{- \omega + \lambda_s ({\bf
q})},
$$
$$
A_s ({\bf q}) = \frac{1}{N} \sum \limits_{\bf k} \sqrt{\Delta G_{\bf k}
({\bf q})} e_{\bf k}^{(s)} ({\bf q}).\eqno(51)
$$
Neglecting in Eq. (36) the operator $\hat M$, we have $e_{\bf k}^{(0)}
({\bf q}) \sim \sqrt {\Delta G_{\bf k} ({\bf q})}$, whence $A_0^2 ({\bf
q}) = - 2 \pi iN(E), \, \, A_s ({\bf q}) = 0, \, \, s \neq 0$. Taking
$\hat M$ into account by perturbation theory, we obtain
$$
A_0^2 ({\bf q}) = \frac{- 2 \pi iN(E)}{1 + B({\bf q})}, \, \, \, \, B({\bf
q}) \sim q^2;
$$
$$
A_s ({\bf q})^2 \sim q^2, \,\,\, s \neq 0\eqno(52)
$$
and comparing the expressions (51) with Eq. (35) gives
$$
D (\omega, {\bf q}) q^2 = i \lambda_0 ({\bf q}) [1 + B ({\bf q})] - i
\omega B ({\bf q}),
$$
$$
\phi_{reg} ({\bf q}) \sim q^2.\eqno(53)
$$
The decomposition into regular and irregular parts is not unique and
admits a "gauge transformation"
$$
{\tilde \phi}_{reg} ({\bf q}) = \phi_{reg} ({\bf q}) - 2 \pi N(E) C({\bf
q}),
$$
$$
{\tilde D} (\omega, {\bf q}) q^2 = \frac{D (\omega, {\bf q}) q^2 + i
\omega C({\bf q}) [- i \omega + D(\omega, {\bf q}) q^2]}{1 + C ({\bf q})
[-i \omega + D (\omega, {\bf q}) q^2]},
$$
$$
C({\bf q}) \sim q^2\eqno(54)
$$
up to which the identity (53) is valid. For this reason, it is convenient
to set by definition
$$
\lambda_0 ({\bf q}) = - iD (\omega, {\bf q}) q^2\eqno(55)
$$
making the assumption that the diffusion coefficient $D (\omega, {\bf q})$
determined in this manner is related to be observed diffusion coefficient
$D_{obs} (\omega, {\bf q})$ by relations of the type (53) and (54). For
any $B({\bf q})$ and $C({\bf q})$, we have $D (0, 0) = D_{obs} (0, 0)$,
and $D (0, {\bf q})$ and $D_{obs} (0, {\bf q})$ vanish simultaneously. In
practice, the difference between $D (\omega, {\bf q})$ and $D_{obs}
(\omega, {\bf q})$ is not important. The point is that the spatial
dispersion of $D (\omega, {\bf q})$ on the scale $q \sim \Lambda$
($\Lambda$ is a parameter of the order of the inverse interatomic
distance) is of little interest; only the "anomalous" dispersion,
determined by the scale $\xi^{-1}$, which can arise near the Anderson
transition, is of interest. The quantity $B ({\bf q})$ does not contain
anomalous dispersion, since it is determined by the function $\Delta
G_{\bf k} ({\bf q})$, which is regular at the transiion point, and the
function $e_{\bf k}^{(0)} ({\bf q})$, which can be assumed to be constant
(Sec. 5.4); this is also true of the quantity $C({\bf q})$, relating,
according to Eq. (54), two regular functions (see, however, Sec. 4). On
the basis of what we have said above, the expression (38) assumes the form
$$
\phi_{{\bf kk}^\prime} ({\bf q}) = \frac{if_{\bf k}^{(0)} ({\bf q})
f_{{\bf k}^\prime}^{(0)} ({\bf q})}{-i \omega + D (\omega, {\bf q}) q^2} +
\phi_{{\bf kk}^\prime}^{(1)} ({\bf q}), \qquad \, \, \, \, \phi_{{\bf
kk}^\prime}^{(1)} ({\bf q}) \sim q^2.\eqno(56)
$$
It follows from the relation (9) that $\phi_{{\bf kk}^\prime} ({\bf q})$
contains a diffusion pole in the limit ${\bf k} + {\bf k}^\prime
\rightarrow 0$, which can be separated from $\phi_{{\bf kk}^\prime}^{(1)}
({\bf q})$:
$$
\phi_{{\bf kk}^\prime} ({\bf q}) = \frac{if_{\bf k}^{(0)} ({\bf q})
f_{{\bf k}^\prime}^{(0)} ({\bf q})}{-i \omega + D (\omega, {\bf q}) q^2} +
\frac{if_{{\bf k}-{\bf k}^\prime+{\bf q}/2}^{(0)} ({\bf k} + {\bf
k}^\prime) f_{{\bf k}^\prime - {\bf k} + {\bf q}/2}^{(0)} ({\bf k} + {\bf
k}^\prime)}{-i \omega + D (\omega, {\bf k} + {\bf k}^\prime)({\bf k} +{\bf
k}^\prime)^2} +
\phi_{{\bf kk}^\prime}^{reg} ({\bf q}).\eqno(57)
$$

In diagrammatic language, the pole in the limit $q \rightarrow 0$ is
related to the fact that for the diagrams containing two or more blocks $U$
(Fig. 2a), the contour of integration in the integrals

\begin{figure}
\centerline{\includegraphics[width=5.1 in]{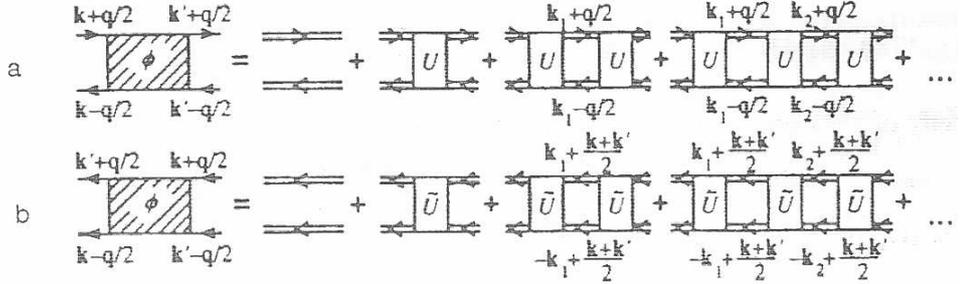}} \caption{
\,\,a
--- Structure of the diagrammatic series for $\phi_{{\bf
kk}^\prime} ({\bf q})$.\,\, b --- same, with the upper $G$ line
inverted; the $U$ and $\tilde U$ blocks are topologically
equivalent, but they correspond to different values of the
momenta.} \label{fig2}
\end{figure}
$$
\int d^d k_i G_{{\bf k}_i + {\bf q}/2}^R G_{{\bf k}_i - {\bf
q}/2}^A\eqno(58)
$$
is confined between the poles of two Green's functions. For small $U$ the
divergence in the expression (58) in the limit $\omega, q \rightarrow 0$
is limited only by the small damping $\mbox{Im} \Sigma$ in the
denominators of the $G$ functions and compensates the smallness associated
with the addition of an extra block $U$; all diagrams in Fig. 2a are found
to be of the same order, and the series diverges, leading to a diffusion
pole. For arbitrary $U$ the divergence of the series in the limit
$\omega, q \rightarrow 0$ is guaranteed by the Ward identity (13). It is
important that the diffusion pole is determined by diagrams with a large
number of blocks $U$. Since $G ({\bf r}_1, {\bf r}_2) = G ({\bf r}_2, {\bf
r}_1)$ the result for $\phi_{{\bf kk}^\prime} ({\bf q})$ will remain
unchanged, if in constructing the diagrams the upper $G$ line is reversed;
then the diagram contains blocks $\tilde U$ (Fig. 2b), topologically
equivalent to the blocks $U$, but taken for other values of the momenta.
Now the poles of the two $G$-functions converge toward one another as
${\bf k} + {\bf k}^\prime \rightarrow 0$, giving a second diffusion pole
in the expression (57). When the upper $G$ line in Fig. 2a is reversed,
the diagrams containing two or more $U$ blocks become irreducible and
enter into a $\tilde U$ block (Fig. 2b) and, conversely, reversion of the
$G$ line in the diagram with one $U$ block generates the entire sequence
of diagrams in Fig. 2b with more than two $\tilde U$ blocks. For this
reason, the second pole term in the expression (57) is contained, with no
changes, in $U_{{\bf kk}^\prime} ({\bf q})$, differing only by the
contribution of the four external $G$ lines. The result (12) with a
function $F ({\bf k}, {\bf k}^\prime, {\bf q})$ of the form
$$
F ({\bf k,k}^\prime, {\bf q}) = if_{({\bf k}-{\bf k}^\prime+{\bf
q)}/2}^{(0)} ({\bf k} + {\bf k}^\prime) f_{({\bf k}^\prime - {\bf k} + {\bf
q)}/2}^{(0)} ({\bf k} + {\bf k}^\prime)
 (G_{{\bf k} + {\bf q}/2}^R G_{{\bf k} - {\bf q}/2}^A G_{{\bf
k}^\prime + {\bf q}/2}^R G_{{\bf k}^\prime - {\bf q}/2}^A )^{-1}\eqno(59)
$$
is valid for $U_{{\bf kk}^\prime} ({\bf q})$. This proves the
Vollhardt--W${\rm {\ddot o}}$lfle hypothesis. \\

\begin{center}
{\bf 4. BEREZINSKII--GOR'KOV CRITERION AND ITS CONSEQUENCES} \\
\end{center}

The spectral density (30) contains a singular contribution $\sim \delta
(\omega)$, originating from terms with $s = s^\prime$, which is finite in
the localized phase and vanishes in the delocalized phase in the
thermodynamic limit. This is the Berezinskii--Gor'kov localization
criterion [24]. The $\delta (\omega)$ singularity in $\langle \rho_E
\rho_{E + \omega} \rangle_{\bf q}$ leads to, by virtue of the Eq. (32), a
$1/ \omega$ singularity in the function $\phi ({\bf q})$ (Ref. 10)
$$
\phi ({\bf q}) = \frac{2 \pi N(E)}{- i \omega} A ({\bf q}) + \phi_{reg}
({\bf q}),\eqno(60)
$$
$$
A ({\bf q}) = \int d {\bf r} e^{-i {\bf q} {\bf r}} A ({\bf r}),
$$
$$
A ({\bf r}) = \frac{1}{N(E)} \langle \sum \limits_s | \psi_s ({\bf r})|^2
| \psi_s (0)|^2 \delta (E - \epsilon_s) \rangle .\eqno(61)
$$
A number of important consequences follow from the relation (60).

{\bf 1.} Comparing Eqs.(60) and (35) shows that in the localized phase $D
(\omega, {\bf q}) \sim \omega$. A slower dependence would destroy the $1/
\omega$ singularity in Eq. (60) and a more rapid dependence would cause
the dependence on $\bf q$ to vanish in the singular part (35); such a
dependence obviously exists according to Eq. (60). This result, valid in
the $D (\omega, {\bf q})$ gauge, in which the functions $\phi_{reg} ({\bf
q})$ in Eqs. (35) and (60) are identical, remains valid in any other gauge
[see Eqs. (53) and (54)]. Therefore
$$
D (\omega, {\bf q}) = (- i \omega) d (q),\eqno(62)
$$
where it is assumed that the limit $\omega \rightarrow 0$ is taken in the
function $d(q)$. Therefore it follows from the Berezinskii--Gor'kov
criterion that $D (0, {\bf q})$ vanishes for all $\bf q$. This completes
the proof of all of the main localization criteria [6, 10]. In view of Eq.
(62), the second diffusion pole in Eq. (57) leads to the singularity $1/
\omega$ in the sum over $s$ in Eq. (51). To eliminate this singularity
from $\phi_{reg} ({\bf q})$ the expression (54) must be transformed with
$C({\bf q}) \sim 1/ \omega$, without destroying  the
proportionality of $D (\omega, {\bf q})$ and ${\tilde D} (\omega, {\bf
q})$ to the frequency. The function $C ({\bf q})$ is determined, by virtue
of Eq. (57), by the quantities $\Delta G_{\bf k} ({\bf q})$ and $e_{\bf
k}^{(0)} ({\bf q})$, which are regular at the transition point and do not
lead to anomalous dispersion, while the associated renormalization of $D
(\omega, {\bf q})$ is small near the transition because of the divergence
of $d(q)$ (see below).

{\bf 2.} A relation between the diffusion coefficient and the properties
of the wave functions follows from Eqs. (35), (60), and (62):
$$
\frac{1}{1 + d(q)q^2} = A ({\bf q}) = \int d {\bf r} e^{-i {\bf q}
\cdot {\bf r}} A ({\bf r}).\eqno(63)
$$
Exponential localization of the wave functions leads to exponential decay
of $A ({\bf r})$ at large $r$ [see Eq. (61)] and
the finiteness of the coefficients in the expansion over $\bf q$ of the
right-hand side of Eq. (63). Because of isotropy in the mean, there are no
odd powers of $\bf q$ and $d(q)$ is a regular function of $q^2$. It is
important that this function does not contain noninteger powers of $q$, which
arise naturally in the case of diffusion over fractal structures [41]. The
reality and positiveness of $d(q)$ follow from the reality of $A ({\bf
q})$  and the inequalities $0 \leq A({\bf q}) \leq 1$ [10, 24].

{\bf 3.} Restrictions on the form of the spatial dispersion of $D (\omega,
{\bf q})$ follow from the relation (63). In the localized phase the
spatial dispersion is determined by the expansion\footnote{In expansions
of the type (64) arbitrary coefficients are assumed. Taking
them into account is beyond accuracy of the present  analysis.}
$$
1 + d(q)q^2 = \xi^{\beta_0} + \xi^{\beta_1} q^2 + \xi^{\beta_2} q^4 + ...
+ \xi^{\beta_n} q^{2n} + ...,
$$
$$
\beta_0 = 0,\eqno(64)
$$
where $\beta_n \geq 0$, since the contributions, associated with the atomic
scale $\Lambda^{-1}$ and corresponding to $\beta_n = 0$, obviously exist.

Different estimates show that the smoothed (over oscillations) behavior of
the squared modulus of a typical wave function has the form
$$
|\psi ({\bf r})|^2 = \mbox{const} \left \{ \begin{array}{cc}
r^{-b}, & \Lambda^{-1} \alt r \alt \xi \\
\mbox{exp} (-r/ \xi), & r \agt \xi
\end{array} \right . .\eqno(65)
$$
This behavior should actually be expected on the basis of Poincare's
theorem on the analytic dependence of the solution of a differential
equation on a parameter. If the behavior of the wave function at the
transition point is characterized by the exponent $b$ ($0 \leq b \leq
\infty$), $|\psi_c ({\bf r})|^2 \sim r^{-b}$, then near the transition
we have $\psi ({\bf r}) \approx \psi_c ({\bf r})$ for
sufficiently small $\bf r$,
as a consequence of Poincare's theorem. The theorem is valid only for a
finite region, whose maximum size is determined by the scale $\xi$ on which
the exponential decrease of $\psi ({\bf r})$ starts. By virtue of Eq. (61)
the function $A ({\bf r})$ has a similar behavior
 $$
A ({\bf r}) = \mbox{const} \left \{ \begin{array}{cc}
r^{-d- \zeta}, & \Lambda^{-1} \alt r \alt \xi \\
\mbox{exp} (-r/ \xi), & r \agt \xi
\end{array} \right . ,\eqno(66)
$$
where const is chosen from the condition $A (q) = 1$ at $q = 0$. The
series expansion of $A (q)$
$$
A ({\bf q}) = \xi^{\gamma_0} + \xi^{\gamma_1} q^2 + \xi^{\gamma_2} q^n +
... + \xi^{\gamma_n} q^{2n} + ..., \, \, \, \, \gamma_0 = 0\eqno(67)
$$
and the estimate of the integrals arising in Eq. (61) show that only two
variants are possible: (a) $\gamma_n = 2n$ for $\zeta < 0$ and (b)
$\gamma_n = \mbox{max} \{ 0, 2n - \zeta \}$ for $\zeta > 0$. Substituting
the expressions (64) and (67) into Eq. (63) gives a relation between
$\gamma_n$ and $\beta_n$
$$
\gamma_n = \mbox{max}_{i+j+k+...=n} \{ \beta_i + \beta_j + \beta_k + ...
\} ,\eqno(68)
$$
leading to the two possibilities for the exponents $\beta_n$: $\beta_1 =
2, \, \, \beta_n \leq 2n$ for $\zeta < 0$ (Fig. 3a)
\begin{figure}
\centerline{\includegraphics[width=5.1 in]{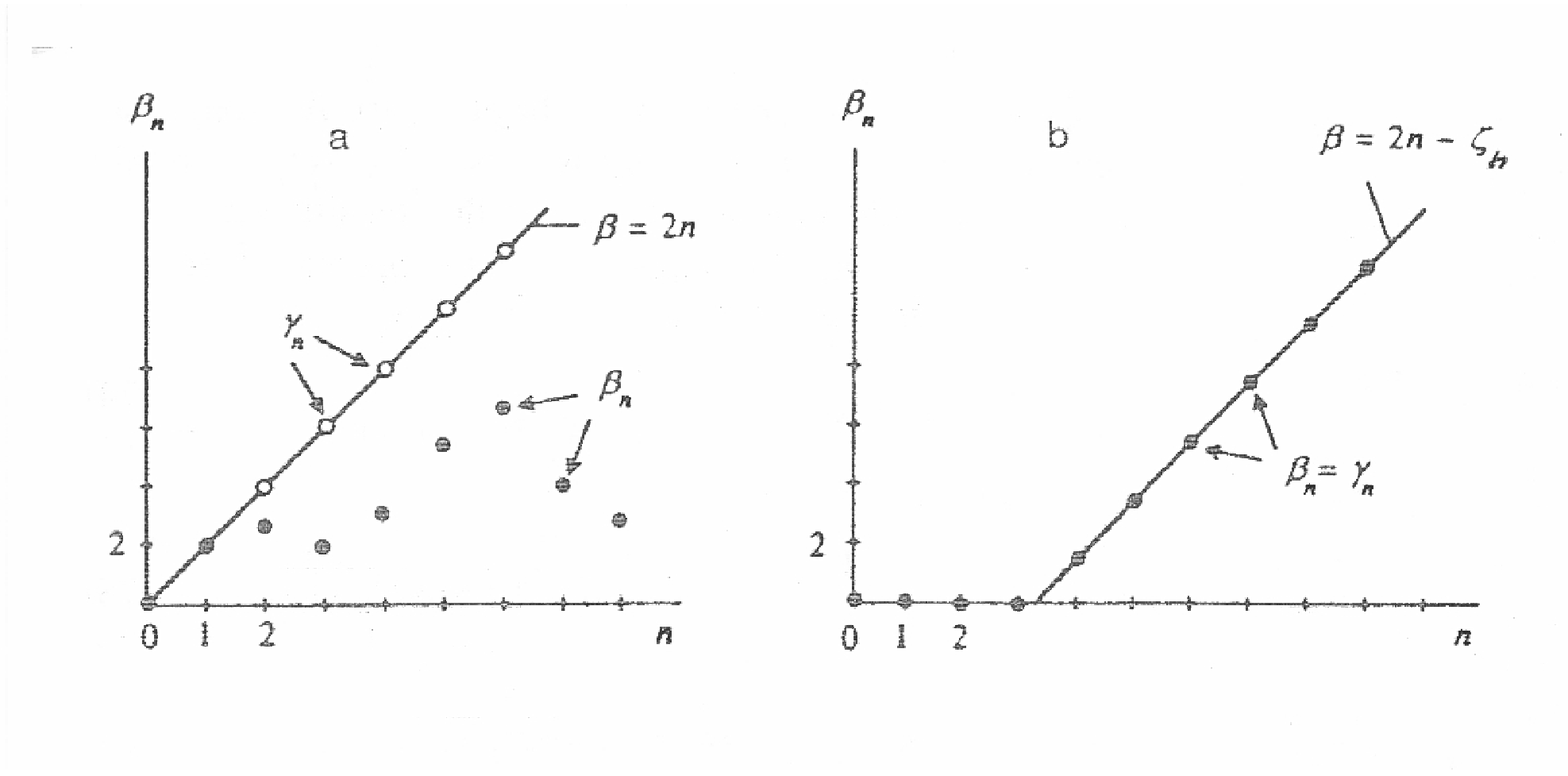}} \caption{
\,\,Possible configurations
of the exponents $\beta_n$ and $\gamma_n$
for $\zeta < 0$ (a) and $\zeta > 0$ (b). } \label{fig3}
\end{figure}
and $\beta_n =
\mbox{max} \{ 0, 2n - \zeta \}$ for $\zeta > 0$ (Fig. 3b). For these
results to be valid it is important only that if the integral of $A ({\bf
r})r^n$ diverges in the limit $\xi \rightarrow \infty$ as $\xi^a$, then
the integral of $A ({\bf r})r^{n+m}$ should diverge as $\xi^{a+m}$, since
it is determined by the region $r \sim \xi$. The specific approximation
(66) is actually not used, but it is convenient for interpreting the
results.

To determine the localization length $\xi$ from the known
diffusion coefficient, in general, it is necessary to know all the
exponents $\beta_n$. The result $D (\omega, 0) \sim (-i \omega) \xi^2$
proposed in Refs. 18 and 10 is valid only for $\zeta < 0$. From Eq. (25)
we obtain for the permittivity
$$
\epsilon (0,0) = 1 + 4 \pi e^2 N (\epsilon_F) d(0) =
\left \{ \begin{array}{cc} \sim \xi^2, & \zeta < 0 \\ \sim \xi^{2- \zeta}, &
0 < \zeta < 2 \\ \sim 1, & \zeta > 2 \end{array} \right .  .\eqno(69)
$$
In the framework of the general analysis, $\epsilon (0, 0)$ can diverge
according to a law that is different from $\xi^2$ (obtained by cutting off
the metallic behavior of $\epsilon (0, {\bf q}) \sim q^{-2}$ on the scale
$q \sim \xi^{-1}$). More than that,  $\epsilon (0, 0)$ can even be finite
in the limit $\xi \rightarrow \infty$ (see the discussion in Refs. 23, 40,
and 41).  \\

\begin{center}
{\bf 5. BASIC STRUCTURE OF THE THEORY} \\
\end{center}

It is convenient to begin the construction of the theory by analyzing the
localized phase, while the metallic state will be obtained as a result of the
instability of the localized phase. \\

\begin{center}
{\bf 5.1. Spectrum of the operator ${\bf {\hat L}}$ in the localized
phase}
\end{center}

Let $M$ be the set of values of the index $s$ that enumerates the
eigenvalues $\lambda_s$ of the operator $\hat L$. We shall show that in
the localized phase the decomposition
$$
M = M_0 \oplus M_1 \oplus M_\infty\eqno(70)
$$
such that
$$
\lambda_s = \left \{ \begin{array}{ccc}
\omega \nu_s, & s \in M_0 & \\
\nu_s, & s \in M_1 \ , & \, \qquad \, \, \, \nu_s \sim 1 \\
\nu_s/ \omega, & s \in M_\infty
\end{array} \right . \eqno(71)
$$
(Fig. 4) is valid.
\begin{figure}
\centerline{\includegraphics[width=5.1 in]{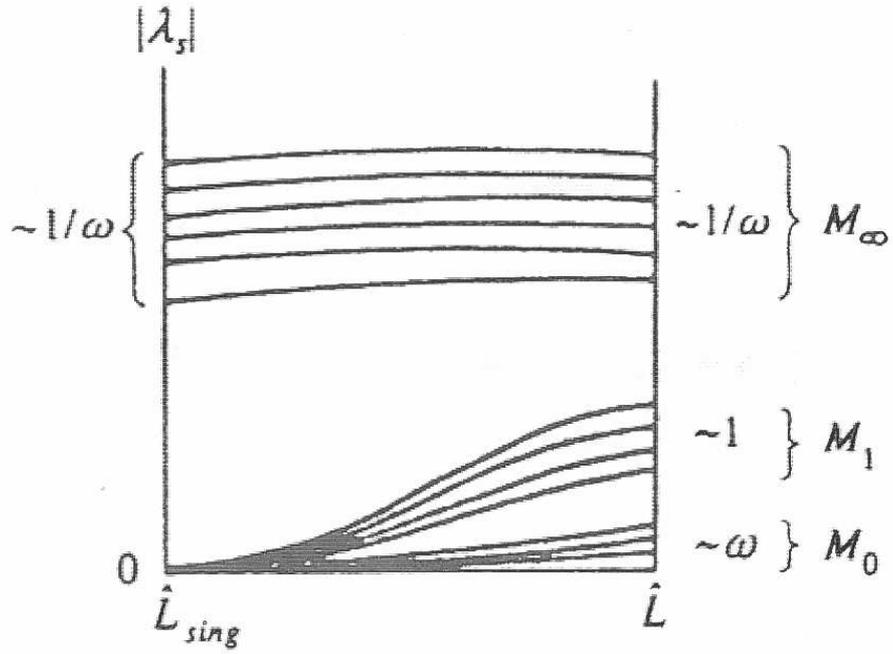}} \caption{
\,\,Evolution of the spectrum of eigenvalues $\lambda_s$ on
transferring from ${\hat L}_{sing}$ to ${\hat L}$, i.e., with the
"gradual switching on" of the operator ${\hat L}_{reg}$. }
\label{fig4}
\end{figure}
The set $M_0$ is not empty, since it contains the
element $\lambda_0 \sim \omega$, related with the diffusion coefficient.
We shall show that it is not the only element. According to Eq. (57),
$\phi_{{\bf kk}^\prime} ({\bf q})$ contains the singularity $ 1/
\omega$, associated with two diffusion poles. In Eq. (38) this singularity
originates from terms with $s \in M_0$. Comparing these two
representations and taking into account the fact that the diffusion pole
at $q = 0$ corresponds to the term with $s = 0$ in Eq. (38), we obtain
$$
\frac{f_{{\bf k}-{\bf k}^\prime+{\bf q}/2}^{(0)} ({\bf k} + {\bf k}^\prime)
f_{{\bf k}^\prime - {\bf k} + {\bf q}/2}^{(0)} ({\bf k} + {\bf
k}^\prime)}{1 + d ({\bf k} + {\bf k}^\prime) ({\bf k} + {\bf k}^\prime)^2}
= i \sum \limits_{s \in M_0^\prime} \frac{f_{\bf k}^{(s)} ({\bf q})
f_{{\bf k}^\prime}^{(s)} ({\bf q})}{1 + \nu_s ({\bf q})},\eqno(72)
$$
where $M_0^\prime$ is the set $M_0$ without the element $s = 0$. Since
$d (q)$ diverges as $\xi \rightarrow \infty$ [see Eq. (64)], the left-hand
side of Eq. (72) contains a $\delta$-function singularity at
${\bf k} + {\bf k}^\prime = 0$, which terms  $f_{\bf
k} f_{{\bf k}^\prime}$ on the right-hand side of Eq. (72) with ${\bf
k}^\prime = - {\bf k}$ cannot have at the arbitrary point
$\bf k$. The same is true for the sum of a finite
number of such terms. The example of the Fourier expansion
$$
\frac{1}{1 + d ({\bf k} + {\bf k}^\prime) ({\bf k} + {\bf k}^\prime)^2}
= \sum \limits_{\bf x} A_{\bf x} e^{i({\bf k} + {\bf k}^\prime) \cdot {\bf
x}} \eqno(73)
$$
shows that the pole term in Eq. (72) can be reproduced by an infinite
number of terms $f_{\bf k} f_{{\bf k}^\prime}$ and that this
does not require a complete system of functions (eliminating from the sum
in Eq. (73) terms with small $\bf x$ leads to the appearance of a smooth
component, but it does not change the singularity at ${\bf k} +
{\bf k}^\prime = 0$). It is clear now,
that the set $M_0$ contains an infinite number of elements, but generally
speaking it does not coincide with the set $M$.

Sadovskii [10, 44] proposed a localization criterion according to which a
nontrivial solution of the homogeneous Bethe--Solpeter equation appears in
the limit $\omega \rightarrow 0$. We can make a stronger assertion:
An infinite number of such solutions appears
at the transition point.

The following decomposition of the operator $\hat L$ follows from Eqs.
(36) and (12):
$$
{\hat L} = {\hat L}_{reg} + {\hat L}_{sing},\eqno(74)
$$
$$
{\hat L}_{reg} \psi_{\bf k} \equiv {\hat M} \psi_{\bf k} + \Delta \Sigma_{\bf
k} ({\bf q}) \psi_{\bf k} -
 \frac{1}{N} \sum \limits_{{\bf k}^\prime}
U_{{\bf kk}^\prime}^{reg} ({\bf q}) \sqrt{\Delta G_{\bf k} ({\bf q})
\Delta G_{{\bf k}^\prime} ({\bf q})} \psi_{{\bf k}^\prime},\eqno(75)
$$
$$
{\hat L}_{sing} \psi_{\bf k} \equiv - \frac{1}{N} \sum \limits_{{\bf
k}^\prime} U_{{\bf kk}^\prime}^{sing} ({\bf q}) \sqrt{\Delta G_{\bf k}
({\bf q}) \Delta G_{{\bf k}^\prime} ({\bf q})} \psi_{{\bf
k}^\prime}.\eqno(76)
$$
In the localized phase the diffusion pole in $U^{sing}$ gives a $1/
\omega$ singularity,
$$
{\hat L} = {\hat L}_{reg} + \frac{{\hat L}_1}{\omega},\eqno(77)
$$
where the limit $\omega \rightarrow 0$ has been taken in the operator
${\hat L}_1$ and terms of higher order in $\omega$ are included in ${\hat
L}_{reg}$. From Eqs. (76), (12), and (72) we obtain the following
representation for ${\hat L}_1$
$$
{\hat L}_1 \psi_{\bf k} = \frac{1}{N} \sum \limits_{{\bf k}^\prime}
\left ( \sum \limits_{s \in M_0^\prime} \frac{g_{\bf k}^{(s)} ({\bf q})
g_{{\bf k}^\prime}^{(s)} ({\bf q})}{1 + \nu_s ({\bf q})} \right )
\psi_{{\bf k}^\prime},
$$
$$
g_{\bf k}^{(s)} ({\bf q}) = \frac{f_{\bf k}^{(s)} ({\bf q}) \sqrt{\Delta
G_{\bf k} ({\bf q})}}{G_{{\bf k} + {\bf q}/2}^R G_{{\bf k} - {\bf
q}/2}^A}.\eqno(78)
$$
It is clear, that the eigenvectors of the operator $\hat L_1$,
corresponding to nonzero eigenvalues, lie in the subspace constructed on
the vectors $g_{\bf k}^{(s)} ({\bf q})$, and the number of eigenvalues is
equal to the number of elements in $M_0^\prime$. The nonzero eigenvalues
of ${\hat L}_1$ correspond to the eigenvalues $\sim 1/ \omega$ of the
operator ${\hat L}_{sing}$.

The overall picture is as follows (Fig. 4). The operator ${\hat L}_{sing}$
has an infinite number of eigenvalues $\sim 1/ \omega$ and an infinite
number of eigenvalues equal to zero. When the operator ${\hat L}_{reg}
\sim 1$ is added, the eigenvalues $\sim 1/ \omega$ change very little and
form the set $M_\infty$ of the operator $\hat L$; the zero eigenvalues
become, generally speaking, of order one, forming the set $M_1$, but
infinite number
of them remain $\sim \omega$ and lie in the set $M_0$. The number of
elements in $M_\infty$ is equal to the number of elements in $M_0^\prime$;
no assertions can be made with respect to the set $M_1$, but this is not
important for what follows. \\

\begin{center}
{\bf 5.2. Relation between ${\bf {\hat L}}$ and ${\bf {\hat L}_{sing}}$}
\end{center}

We now introduce the spectral representation for the singular part of the
operator ${\hat L}_{sing} = {\hat L}_1/ \omega$
$$
{\hat L}_1 = \sum \limits_s |u_s \rangle \eta_s \langle u_s|, \, \qquad \, \,
\, \eta_s = \left \{ \begin{array}{cc} 0 & s \in M_0 \oplus M_1 \\ \sim 1 & s
\in M_\infty \end{array} \right .\eqno(79)
$$
and find a relation between $\hat L$ and ${\hat L}_{sing}$, regarding
${\hat L}_{reg}$ as a perturbation. For $s \in M_\infty$ the ordinary
perturbation theory can be used, since all differences of the eigenvalues
$\sim 1/ \omega$ and a regular expansion in powers of $\omega$ is
obtained:
$$
|e_s \rangle = |u_s \rangle + \omega \sum \limits_{s^\prime \neq s}
\frac{\langle u_{s^\prime} |{\hat L}_{reg} |u_s \rangle}{\eta_s -
\eta_{s^\prime}} |u_{s^\prime} \rangle,
$$
$$
\lambda_s = \frac{\eta_s}{\omega} + \langle u_s| {\hat L}_{reg} |u_s
\rangle, \, \, \, \, \, \, \, \, \, \, \,  s \in M_\infty .\eqno(80)
$$
For $s \in M_0 \oplus M_1$ we seek the eigenfunctions of $\hat L$ in the
form
$$
|e \rangle = \sum \limits_{s \in M_0 \oplus M_1} C_s|u_s \rangle + \omega
\sum \limits_{s \in M_\infty} D_s|u_s \rangle ,\eqno(81)
$$
where $C_s, \, D_s \sim 1$. Substituting (81) in the eigenvalue
equation, we obtain a system of equations for $C_s$ and $D_s$, which can
be solved by iterations in $\omega$. Eliminating $D_s$, we obtain to first
order in $\omega$
$$
\sum \limits_{s^\prime \in M_0 \oplus M_1} (\lambda_s \delta_{ss^\prime}
- T_{ss^\prime}) C_{s^\prime} = 0, \qquad \, \, \, s \in M_0 \oplus M_1,
$$
$$
T_{ss^\prime} = \langle u_s| {\hat L}_{reg} |u_{s^\prime} \rangle - \omega
\sum \limits_{s^{\prime \prime} \in M_\infty} \frac{\langle u_s| {\hat
L}_{reg} |u_{s^{\prime \prime}} \rangle \langle u_{s^{\prime \prime}} |
{\hat L}_{reg} |u_{s^\prime} \rangle}{\eta_{s^{\prime \prime}}},\eqno(82)
$$
i.e., an ordinary secular equation taking into account the first
correction from transitions into states with $s \in M_\infty$. \\

\begin{center}
{\bf 5.3. Mechanism for satisfying the Ward identity}
\end{center}

We now demonstrate the cancellation of the singular contribution $\sim 1/
\omega$, associated with the diffusion pole in $U_{{\bf kk}^\prime} ({\bf
q})$, on the right-hand side of the Ward identity (13). The specific form
of ${\hat L}_{reg}$ was not used in Secs. 5.1 and 5.2. To determine
${\hat L}_{reg}$ in the form (75) with ${\hat M} \equiv 0$ we have
$$
\frac{1}{N} \sum \limits_{\bf k} \sqrt{\Delta G_{\bf k} ({\bf q})} e_{\bf
k}^{(s)} ({\bf q}) = 0, \, \, \, \qquad \, s \in M_\infty,\eqno(83)
$$
since $\sqrt{\Delta G_{\bf k} ({\bf q})}$ is the exact eigenfunction of
$\hat L$ belonging to the set $M_0$. By virtue of the relation (80), the
difference of $| u_s \rangle$ from $| e_s \rangle$ for $s \in M_\infty$ is
of order $\omega$ for any ${\hat L}_{reg}$, whence
$$
\frac{1}{N} \sum \limits_{\bf k} \sqrt{\Delta G_{\bf k} ({\bf q})} u_{\bf
k}^{(s)} ({\bf q}) = O (\omega), \, \, \qquad \, \, s \in M_\infty.\eqno(84)
$$
Comparing Eqs. (76) and (79), we have
$$
- U_{{\bf kk}^\prime}^{sing} ({\bf q}) \sqrt{\Delta G_{\bf k} ({\bf q})
\Delta G_{{\bf k}^\prime} ({\bf q})} = \frac{1}{\omega} \sum \limits_{s
\in M_\infty} u_{\bf k}^{(s)}
({\bf q}) \eta_s ({\bf q}) u_{{\bf k}^\prime} ^{(s)} ({\bf q})
,\eqno(85)
$$
so that the singular contribution on the right-hand side of Eq. (13),
taking into account Eq. (84), has the form
$$
\frac{1}{N} \sum \limits_{{\bf k}^\prime} U_{{\bf kk}^\prime}^{sing}
\Delta G_{{\bf k}^\prime} ({\bf q}) = - \frac{1}{\omega} \sum \limits_{s
\in M_\infty} \frac{u_{\bf k}^{(s)} ({\bf q}) \eta_s ({\bf
q})}{\sqrt{\Delta G_{\bf k} ({\bf q})}} \frac{1}{N}
 \sum \limits_{{\bf k}^\prime} \sqrt{\Delta G_{{\bf k}^\prime} ({\bf
q})} u_{{\bf k}^\prime}^{(s)} ({\bf q})  = \frac{O
(\omega)}{\omega}\eqno(86)
$$
and the $1/ \omega$ singularity cancels. For the same reason, there will
be no singularities on the right-hand side of Eq. (13) as the transition
into the metallic phase is approached. In this case the spectrum of the
operator $\hat L$ has the same structure (Fig. 4) with $\omega$ replaced
by $D_0$ --- the characteristic value of the diffusion coefficient (Sec.
6.2) --- and we obtain $O(D_0)/D_0$ on the right-hand side of Eq. (86). \\

\begin{center}
{\bf 5.4. Symmetry approach}
\end{center}

The symmetry of the system is clearly expressed in the properties of the
operator ${\hat L}_{sing}$:

(a) The spatial uniformity in the mean makes it possible to introduce the
three-momentum notations (Fig. 1d) and to introduce the operator $\hat L$
in general and the operator ${\hat L}_{sing}$ in particular.

(b) The isotropy in the mean, combined with time-reversal invariance,
guarantees that $\hat L$ and ${\hat L}_{sing}$ are symmetrical and the
existence of orthonormal bases of eigenvectors for them.

(c) As a result of time-reversal invariance, ${\hat L}_{sing}$ has a high
symmetry, manifested in the existence of an infinite number of zero
modes\footnote{Their presence (see Sec.5.1) is associated with the
existence of a diffusion pole for ${\bf k} + {\bf k}^\prime \rightarrow
0$, which follows from Eq. (9) (see Sec. 3). }.

The decomposition (77) represents the operator $\hat L$ as a sum of the
operator ${\hat L}_{sing}$ with a high degree of symmetry and a regular
operator ${\hat L}_{reg}$ of a general form. It is similar to the
decomposition (1) and is convenient for symmetry analysis. The condition
on the transition point will be determined below and the origin of the
parameter $\tau$ will thereby be determined.

Following Sec. 1, we consider the response of the system to a perturbation
$\delta {\hat L}_{reg}$ of a general form. Not all changes in the system
will be important. We decompose the change in the operator $\hat L$ into
two parts
$$
\delta {\hat L} = \delta {\hat L}_\lambda + \delta {\hat L}_e,\eqno(87)
$$
where $\delta {\hat L}_\lambda$ changes the eigenvalues and $\delta {\hat
L}_e$ changes the eigenfunctions of ${\hat L}$. For an infinitesimal
 $\delta {\hat L}$ such a decomposition is trivial --- $\delta {\hat
L}_\lambda$ and $\delta {\hat L}_e$ are the diagonal and off-diagonal
parts of the operator $\delta {\hat L}$ in the representation of the
eigenvectors $| e_s \rangle$. Changes of the $\delta {\hat L}_e$ type do not
change the eigenvalues of $\hat L$ and therefore the diffusion coefficient $D
(\omega, {\bf q})$, directly related with $\lambda_0 ({\bf q})$. The
diffusion coefficient determines uniquely the location of the system ---
in a localized phase, in a metallic state, or at the transition point. It is
clear that the changes $\delta {\hat L}_e$ do not drive the system out of the
transition point, they only displace the system along the critical surface
[1]. Such displacements do not lead to nonanalyticity of the physical
quantities\footnote{Singularities associated with a change in the type of
phase transition --- for example, a second-order phase transition into a
first-order phase transition --- can occur on the critical surface. We assume
that the system is far away from such singularities.} and they can be
ignored. The critical exponents obtained by motion along the normal to the
critical surface are identical to the exponents obtained under an
arbitrary nonzero angle to the tangent plane. Similarly, in perturbations
of the $\delta {\hat L}_\lambda$ type, the part corresponding to a change
in $\lambda_s$ with $s \in M_1 \oplus M_\infty$ need not be considered.

Only the changes in the eigenvalues $\lambda_s$ from the set $M_0$  are
important, and their response to a perturbation is indeed nontrivial. Let
the system lie deep in the localized phase. A small perturbation $\delta
{\hat L}_{reg}$ does not drive the system out of the state of localization
and preserves the proportionality $\lambda_s \sim \omega$ for $s \in M_0$. On
the other hand, a perturbation $\delta {\hat L}_{reg}$ of a general form
possesses nonzero matrix elements with the respect to the eigenvectors
$| e_s \rangle$ of the subspace $M_0$ and should lead to small but
nonvanishing, in the limit $\omega \rightarrow 0$, values of $\lambda_s$.
The resolution of this contradiction will lead to the self-consistency
equation (Sec. 5.6). \\

\begin{center}
{\bf 5.5. "Rotation" of the singular operator}
\end{center}

To formulate an adequate language for the further discussion, we shall
consider the following problem of the "rotation" of a singular operator.

Let the decomposition (77), where $\omega \rightarrow 0$, be valid for the
operator $\hat L$. The operator ${\hat L}_{reg}$ acts in the space
$\Omega$, while the operator ${\hat L}_1$ has nonzero eigenvalues $\sim 1$
in the subspace $\Omega_1$, which is a part of $\Omega$, $\Omega = \Omega_0
\oplus \Omega_1$. This justifies retaining in Eq. (77) two terms of different
orders. Let $\delta {\hat L}_1$ be a perturbation of the operator ${\hat
L}_1$. If this perturbation is of a general form, then the addition
$\delta {\hat L}_1/ \omega$ to the operator ${\hat L}$ can be studied by
the standard perturbation theory and gives corrections $\sim 1/ \omega$.
Let the perturbation $\delta {\hat L}_1$ be such, however, that the
operator ${\hat L}_1 + \delta {\hat L}_1$ has the same properties as the
initial operator ${\hat L}_1$. Then the dimension of the subspace
$\Omega_1$ remains the same and only a "rotation" of the operator $L_1$
occurs; in this case $\delta {\hat L}_1$ has no nonzero matrix elements in
$\Omega_0$. What is the result of such a perturbation in the subspace
$\Omega_0$?

Let ${\bar \eta}_s$ and $| {\bar u}_s \rangle$ be the eigenvalues and
eigenvectors of the initial operator ${\hat L}_1$. The operator ${\hat
L}_{reg}$ can be neglected in the "upper" subspace $\Omega_1$, and in the
"lower" subspace $\Omega_0$ a secular equation
in terms of the matrix elements
$\langle {\bar u}_s | {\hat L}_{reg} | {\bar u}_{s^\prime} \rangle$ should
be written out. The perturbation $\delta {\hat L}_1$ produces the change
$\delta u_s \sim \delta {\hat L}_1$ of the eigenvectors $| u_s \rangle$
and the matrix of the secular equation is determined by the elements
$$
\langle u_s | {\hat L}_{reg} |u_{s^\prime} \rangle = \langle {\bar u}_s +
\delta u_s| {\hat L}_{reg}| {\bar u}_{s^\prime} + \delta u_{s^\prime}
\rangle \equiv
 \langle {\bar u}_s| {\hat L}_{reg} + \delta {\hat V}|{\bar
u}_{s^\prime} \rangle .\eqno(88)
$$
The qualitative result is that a limitation of the form of the operator
$\delta {\hat L}_1$ weakens its action on the lower subspace: The
effective perturbation $\delta {\hat V}$ appears to be  $\sim \delta {\hat
L}_1$ instead of $\delta {\hat L}_1 / \omega$ for the operator of general
form.

The change in $| u_s \rangle$ in the subspace $\Omega_1$ can be calculated
by the standard perturbation theory, since all differences of the
eigenvalues $\sim 1$ and a series in the small parameter arises:
$$
|u_s \rangle = |{\bar u}_s \rangle + \sum \limits_{s^\prime \neq s}
\frac{\langle {\bar u}_{s^\prime}| \delta {\hat L}_1| {\bar u}_s
\rangle}{{\bar \eta}_s - {\bar \eta}_{s^\prime}} |{\bar u}_{s^\prime}
\rangle , \, \, \, \, \qquad \, \, s \in \Omega_1 .\eqno(89)
$$
An arbitrary choice of $| u_s \rangle$ that is compatible with the
orthogonality relations can be made, in view of degeneracy, in the
subspace $\Omega_0$. To first order in $\delta {\hat L}_1$ we can set
$$
|u_s \rangle = |{\bar u}_s \rangle - \sum \limits_{s^\prime \in \Omega_1}
\frac{\langle {\bar u}_s | \delta {\hat L}_1| {\bar u}_{s^\prime}
\rangle}{{\bar \eta}_{s^\prime}} |{\bar u}_{s^\prime} \rangle , \, \, \,
\, \, \qquad \, s \in \Omega_0.\eqno(90)
$$
Substituting the expression (90) into Eq. (88), we obtain for the matrix
elements of the effective perturbation
$$
\langle {\bar  u}_s | \delta {\hat V} | {\bar u}_{s^\prime} \rangle = -
\sum \limits_{s^{\prime \prime} \in \Omega_1} \frac{\langle {\bar u}_s |
\delta {\hat L}_1 | {\bar u}_{s^{\prime \prime}} \rangle \langle {\bar
u}_{s^{\prime \prime}} | {\hat L}_{reg} | {\bar u}_{s^\prime} \rangle +
\langle {\bar u}_{s^\prime} | \delta {\hat L}_1 | {\bar u}_{s^{\prime \prime}}
\rangle \langle {\bar u}_{s^{\prime \prime}} | {\hat L}_{reg} | {\bar u}_s
\rangle}{{\bar \eta}_{s^{\prime \prime}}}.\eqno(91)
$$
\\

\begin{center}
{\bf 5.6. Self-consistency equation}
\end{center}

It is now easy to understand how to resolve the contradiction stated in
Sec. 5.4. The perturbation $\delta {\hat L}_{reg}$ produces the change
$\delta d(q)$ in the diffusion coefficient (62), which in view of the
relation
$$
{\hat L}_{sing} \psi_{\bf k} = \frac{1}{N} \sum \limits_{{\bf k}^\prime}
\frac{W ({\bf k}, {\bf k}^\prime, {\bf q}) \psi_{{\bf k}^\prime}}{- i
\omega + D (\omega, {\bf k} + {\bf k}^\prime)({\bf k} + {\bf k}^\prime)^2}
=
$$
$$
= \frac{1}{(-i \omega)} \int \frac{d^d {\tilde q}}{(2 \pi)^d} \, \frac{W
({\bf k}, - {\bf k} + {\tilde {\bf q}}, {\bf q}) \psi_{- {\bf k} + {\tilde
{\bf q}}}}{1 + d ({\tilde q}) {\tilde q}^2} \equiv
 \frac{{\hat L}_1 \psi_{\bf k}}{\omega}\eqno(92)
$$
gives the following change in ${\hat L}_1$:
$$
\delta {\hat L}_1 \psi_{\bf k} = (-i) \int \frac{d^d {\tilde q}}{(2
\pi)^d} \, \frac{{\tilde q}^2 \delta d ({\tilde q})}{[1 + d ({\tilde q})
{\tilde q}^2]^2}
 W ({\bf k}, - {\bf k} + {\tilde {\bf q}}, {\bf q}) \psi_{- {\bf k}
+ {\tilde {\bf q}}}.\eqno(93)
$$
Rotation of the subspace $M_\infty$ of the operator ${\hat L}_{sing}$
produces in the subspace $M_0$ the effective perturbation $\delta {\hat
V}$, which in zeroth order in $\omega$ compensates $\delta {\hat
L}_{reg}$.

Introducing into Eq. (82) the small changes $\delta {\hat L}_{reg}$ and
$\delta {\hat L}_1$ [the latter enters via the change in the
eigenfunctions (90)], we obtain for the matrix of the secular equation
$$
T_{ss^\prime} = \langle {\bar u}_s | {\bar T} + \delta {\hat L}_{reg} +
\delta {\hat V} | {\bar u}_{s^\prime} \rangle \, \, \qquad \, \, \, \,
s,s^\prime \in M_0 \oplus M_1,\eqno(94)
$$
where the overbar denotes the unperturbed value, and $\bar T$ and $\delta
{\hat V}$ are determined by the expressions (82) and (91) (with the
substitution $\Omega_1 \rightarrow M_\infty$, $\Omega_0 \rightarrow M_0
\oplus M_1$). In the terms $\sim \omega$ we confine ourselves to zeroth
order in the increments. The choice of the vectors $| {\bar u}_s \rangle$
in the subspace $M_0 \oplus M_1$ is arbitrary in view of the degeneracy.
We choose them so as to diagonalize the matrix ${ {\bar T}}$ --- then, to
zeroth order in $\omega$, they are identical to the eigenvectors $| {\bar
e}_s \rangle$ of the operator ${\hat {\bar L}}$ [see Eq. (81)]. Since the
eigenvalues of the matrix $ T$ are identical to the eigenvalues of
${\hat L}$, we have
$$
T_{ss^\prime} = {\bar \lambda}_s \delta_{ss^\prime} + \langle {\bar e}_s |
\delta {\hat L}_{reg} + \delta {\hat V} | {\bar e}_{s^\prime} \rangle , \,
\qquad \, \, \, \, \, s,s^\prime \in M_0 \oplus M_1.\eqno(95)
$$
For infinitesimal $\delta {\hat L}_{reg}$ and $\delta {\hat V}$, the
diagonal elements of the matrix $\hat T$ determine the eigenvalues of the
operator ${\hat L}$
$$
\lambda_s = {\bar \lambda}_s + \langle {\bar e}_s | \delta {\hat L}_{reg} +
\delta {\hat V} | {\bar e}_s \rangle , \qquad \, \, \, \, \, \, s \in M_0
\oplus M_1\eqno(96)
$$
and the off-diagonal elements determine the corrections to its
eigenfunctions; the latter correspond to the perturbations of the type
$\delta {\hat L}_e$ (Sec. 5.4) and can be dropped. For constant $| e_s
\rangle$ it is possible to switch in Eq. (96) from infinitesimals to
finite increments. Further, the changes in $\lambda_s$ in the subspace
$M_1$ can be ignored (Sec. 5.4). Finally, we note that fixing $\lambda_0
({\bf q})$ for all $\bf q$ means fixing the diffusion coefficient, which
in turn determines all $\lambda_s ({\bf q})$ with $s \in M_0^\prime$,
which can be reconstructed according to the binary decomposition (72).
Therefore if Eq. (96) is satisfied for $s = 0$
$$
-i [D (\omega, q) q^2 - {\bar D} (\omega , q)q^2] = \langle {\bar e}_0 |
\delta {\hat L}_{reg}| {\bar e}_0 \rangle -
 2 \sum \limits_{s^{\prime \prime} \in
M_\infty} \frac{\langle {\bar e}_0 | \delta {\hat L}_1 | {\bar e}_{s^{\prime
\prime}} \rangle \langle {\bar e}_{s^{\prime \prime}} | {\hat L}_{reg} |{\bar
e}_0 \rangle}{{\bar \eta}_{s^{\prime \prime}}}\eqno(97)
$$
(Eqs. (55), (91), and (80) were employed), then it is automatically
satisfied for all $s$ from $M_0$. It is easy to show (see Appendix) that
the expansion in $\bf q$ of the right-hand side of Eq. (97) contains only
even powers of $\bf q$, and terms $\sim q^0$ are absent in each of the two
terms. Setting
$$
\langle {\bar e}_0 | \delta {\hat L}_{reg}| {\bar e}_0 \rangle = - i q^2
\delta f (q)\eqno(98)
$$
and substituting the expression (93) into Eq. (97), we obtain
$$
D (\omega, q) - {\bar D} (\omega, q) = \delta f (q) + {\hat Q} \delta d
(q) ,\eqno(99)
$$
$$
{\hat Q} \delta d (q) = \int \frac{d^d {\tilde q}}{(2 \pi)^d} \, \frac{B
(q, {\tilde q}) {\tilde q}^2}{[1 + d ({\tilde q}) {\tilde q}^2]^2} \delta d
({\tilde q}),\eqno(100)
$$
where $\delta f(q)$ and $B (q, {\tilde q})$ are regular functions of
general form of the arguments $q^2$ and ${\tilde q}^2$. The quantity
$W({\bf k, k}^\prime, {\bf q})$ in Eq. (92) can be expressed, by virtue of
Eqs. (76), (12), and (59), in terms of functions which are regular at the
transition point. This makes it possible to vary only $d(q)$ on switching
from Eq. (92) to Eq. (93). The equation (99) contains the diffusion
coefficient on the right- and left-hand sides and replaces the
self-consistency equation (16) of the Vollhardt--W${\rm {\ddot o}}$lfle
theory. \\

\begin{center}
{\bf 5.7. Condition on the transition point}
\end{center}

In deep of the localized phase $D (\omega, q)$ and ${\bar D} (\omega, q)$
vanish at $\omega = 0$ and Eq. (99) determines the change $\delta d(q) =
-{\hat Q}^{-1} \delta f (q)$ for the prescribed perturbation $\delta {\hat
L}_{reg}$.  Making the small changes $\delta {\hat L}_{reg}$, we obtain
the corresponding changes $\delta d(q)$, which preserve the
proportionality $D (\omega, q) \sim (- i \omega)$. This situation remains
as long as there exists an operator inverse to $\hat Q$, i.e. as long as
all eigenvalues of $\hat Q$ are nonzero. Let a nonzero eigenvalue of the
operator $\hat Q$ appear at some point in the course of the motion from
the interior of the localized phase. As we shall see below, such a point
corresponds to the physical notions of the Anderson transition.

The divergence of $d (q)$ in the limit $\xi \rightarrow \infty$ (Sec. 4)
means [see Eq. (100)] that at the transition point the operator $\hat Q$
vanishes entirely or on some subspace. For this reason, it should be kept
in mind in the analysis that many or even all eigenvalues $\mu_s$ of the
operator $\hat Q$ can vanish simultaneously at the transition point. It is
 convenient to introduce the critical exponent
$\delta_s \geq 0$  for each of them:
$$
{\hat Q} \phi_s (q) = \mu_s \phi_s (q), \ \qquad  \ \mu_s \sim
\tau^{\delta_s}\eqno(101)
$$

 As $d (q) \rightarrow
\infty$, the changes of the function $B (q, {\tilde q})$ cannot make the
operator $\hat Q$ finite and therefore they do not drive the system out of
the critical point; they only displace it along the critical surface and
can be ignored. So the function $B (q, {\tilde q})$ is considered
as independent of $\tau$. In this case, the equality
$$
\int \frac{d^d q}{(2 \pi)^d} B (q, {\tilde q}) \phi (q) = 0 \eqno(102)
$$
cannot be satisfied for any function $\phi (q)$. Indeed, it corresponds
to the presence of a zero mode for the transposed operator ${\hat Q}^T$ (and
therefore for the operator $\hat Q$ itself) not only at the transition
point but also in a finite interval around it.  \\

\begin{center}
{\bf 6. SOLUTION OF THE SELF-CONSISTENCY EQUATION} \\
\end{center}

\begin{center}
{\bf 6.1. Classification of the possible solutions}
\end{center}

The self-consistency equation for the metallic phase can be derived only by
making specific assumptions about the functional form $D (\omega, q)$. For
this reason, it is convenient to examine several cases which exhaust all
possibilities.

(a) Let there be among the exponents $\delta_s$ in Eq. (101) a maximum
exponent (for definiteness, $\delta_0$), i.e., among the set of soft
modes, one mode is the softest. Then, as the transition is approached, the
component const $\phi_0 (q)$, contained in $\delta f(q)$, will give rise
to an anomalously large response $\mbox{const} \tau^{- \delta_0} \phi_0
(q)$ in the function $\delta d (q)$. For this reason, near the transition
the solution can be sought in the form
$$
D (\omega, q) = D_0 [ \phi_0 (q) + \varphi (q)], \qquad \, \, \varphi (q)
\ll \phi_0 (q).\eqno(103)
$$
For $\phi_0 (q)$ to dominate for all values of $q$, it is necessary that
$\phi_0 (0) \neq 0$, which we shall assume is the case.

(b) Let several exponents have the maximum value $\delta_0 = \delta_1 =
... = \delta_p$, and let at least one of the functions $\phi_0 (q), \,
\phi_1 (q), \, ... \, \phi_p (q)$ [for example $\phi_0 (q)$] be different
from zero at $q = 0$. Then near the transition
$$
D (\omega, q) = D_0 [ \phi_0 (q) + C_1 \phi_1 (q) + ... +
$$
$$
+ C_p \phi_p (q) + \varphi (q)],\eqno(104)
$$
where $C_1 \sim C_2 \sim ... \sim C_p \sim 1, \, \, \varphi (q) \ll
\phi_0 (q)$.

(c) If for  two eigenfunctions  we have $\phi_0 (q)\sim q^{2n_0}, \, \,
\phi_1 (q) \sim q^{2n_1}$ in the limit $q \rightarrow 0$, and $n_0 > n_1, \,
\, \delta_0 > \delta_1$, then in the expansion of $D(\omega, q)$ in $\phi_s
(q)$ both functions must be retained. Although the coefficient of $\phi_0
(q)$ grows more rapidly near a transition, the function $\phi_1 (q)$
dominates for small values of $q$. In the general case, $d (q)$ must be
sought in the form of the expansion (64) with arbitrary $\beta_n$.

Actually, as we shall see below, the case (b) is realized (Sec. 6.3). But
the analysis of this case is virtually identical to the simpler case (a)
(Sec. 6.2), which reproduces the solution of the self-consistent theory of
localization [18]. Analysis of the case (c) requires a special
mathematical apparatus (Sec. 7), and there is no need to extend this
analysis  for the metallic phase,
since the solutions  different from (b) do not exist. \\

\begin{center}
{\bf 6.2. Case of a single dominant mode}
\end{center}

We seek the solution in the form (103). The definition of the operators
${\hat L}_1$ and $\hat Q$ in Sec. 5 presumed that only the dependence on
$\omega$ in the localized phase is investigated. To investigate the
dependence on $\omega$ and $\tau$ it is necessary to take into account the
fact that near the transition and in the metallic phase the magnitude of
the diffusion denominator is determined by the parameter $D_0 \gg \omega$.
Making the decomposition
$$
\frac{1}{-i \omega + D (\omega, q)q^2} = \frac{1}{D_0} \left \{
\frac{1}{\phi_0 (q)q^2}  -
 \frac{\varphi (q) - i \omega/D_0 q^2}
{\phi_0 (q) \left [ - i
\omega/D_0 + \phi_0 (q)q^2 + \varphi (q)q^2 \right ]} \right \}\eqno(105)
$$
we write ${\hat L}_{sing}$ in the form
$$
{\hat L}_{sing} = \frac{{\hat {\bar L}}_1 + \delta {\hat
L}_1}{D_0},\eqno(106)
$$
where ${\hat {\bar L}}_1$ and $\delta {\hat L}_1$ correspond to the first
and second terms in the braces in Eq. (105). Substituting $\delta {\hat
L}_1$ into Eq. (97) gives, instead of Eq. (99), the equation
$$
D (\omega, q) = \tau f (q) + {\hat Q}_R \varphi (q) - \frac{i \omega}{D_0}
{\hat Q}_R q^{-2},\eqno(107)
$$
$$
{\hat Q}_R \psi (q) \equiv \int \frac{d^d \tilde  q}{(2 \pi)^d} \frac{B (q,
{\tilde q})}{\phi_0 ({\tilde q})[- i \omega/D_0 + \phi_0 ({\tilde q}) {\tilde
q}^2]} \psi ({\tilde q}),\eqno(108)
$$
where we have neglected $\varphi (q)$ in the denominator of Eq. (105), and
have written $\delta f (q)$ in the form $\tau f (q)$. We also
accept that the operator ${\hat {\bar L}}_1$ corresponds to the
limit $\omega, \tau \rightarrow 0$ (see Fig. 5 below) and
put ${\bar D} (\omega, q) \equiv 0$. For the decomposition (103)
to be unique, we require
that $\varphi (q)$ satisfy the condition
$$
({\bar \phi}_0 (q), \varphi (q)) = 0\eqno(109)
$$
 expressing the requirement that
$\varphi (q)$ "not contain in itself" the component $\mbox{const} \cdot
\phi_0 (q)$ (${\bar \phi}_0 (q)$ is an eigenfunction of ${\hat Q}_R^T$ that
corresponds to the eigenvalue $\mu_0^R$). Forming the scalar
product of the expression (107) with
${\bar \phi}_0 (q)$, we obtain
$$
D_0 ({\bar \phi}_0, \phi_0) = \tau ({\bar \phi}_0, f) - \frac{i
\omega}{D_0} ({\bar \phi}_0, {\hat Q}_R q^{-2}),\eqno(110)
$$
where the last term is different from zero in view of the impossibility of
Eq. (102), and the term with $\varphi (q)$ is absent since $({\bar
\phi}_0, {\hat Q}_R \varphi) = ( \varphi, {\hat Q}_R^T {\bar \phi}_0) =
\mu_0^R (\varphi, {\bar \phi}_0) = 0$. Written out in detail, Eq. (110) has
the structure
$$
D_0 = A \tau - \frac{i \omega}{D_0} \int \frac{d^d {\tilde q}}{(2 \pi)^d}
\frac{B ({\tilde q})}{\phi_0 ({\tilde q}) {\tilde q}^2 [- i
\omega/D_0 + \phi_0 ({\tilde q})
{\tilde q}^2]} .\eqno(111)
$$
For $d > 4$ the integral is determined by large values of $\tilde q$, and
$-i \omega /D_0$ in the denominator can be neglected. For $d< 4$ the
integral is determined by small $\tilde q$, and we can set ${\tilde q} = 0$
in the slowly varying functions $B ({\tilde q})$ and $\phi_0 ({\tilde
q})$; the region of integration can be taken
infinite and the integral can be made dimensionless. The result for
both cases can be written in the unigue form
$$
D_0 = A \tau + B \left ( - \frac{i \omega}{D_0} \right )^{1/2
\nu},\eqno(112)
$$
introducing exponent $\nu$ according to Eq. (18). The equation (112) has
two types of solutions: in the metallic phase $D_0 = \mbox{const} \neq 0$
as $\omega \rightarrow 0$ and Eq. (112) gives $D_0 = A \tau$ in accordance
with the value $s = 1$ for the conductivity exponent (18); in the
dielectric phase $D_0 = (-i \omega ) \xi^2$  and
$\xi \sim \tau^{- \nu}$ in accordance with the definition of the exponent
of the localization length (in the case at hand, the
configuration of exponents $\beta_n$ corresponds to the case Fig. 3a).
The equation (112) and the values of the
indices $s$ and $\nu$ are identical to those obtained in Ref. [18].

For the case $d > 4$ Eq. (112) reduces to a quadratic equation and it is
easy to trace how the solutions are selected (Fig. 5).
\begin{figure}
\centerline{\includegraphics[width=5.1 in]{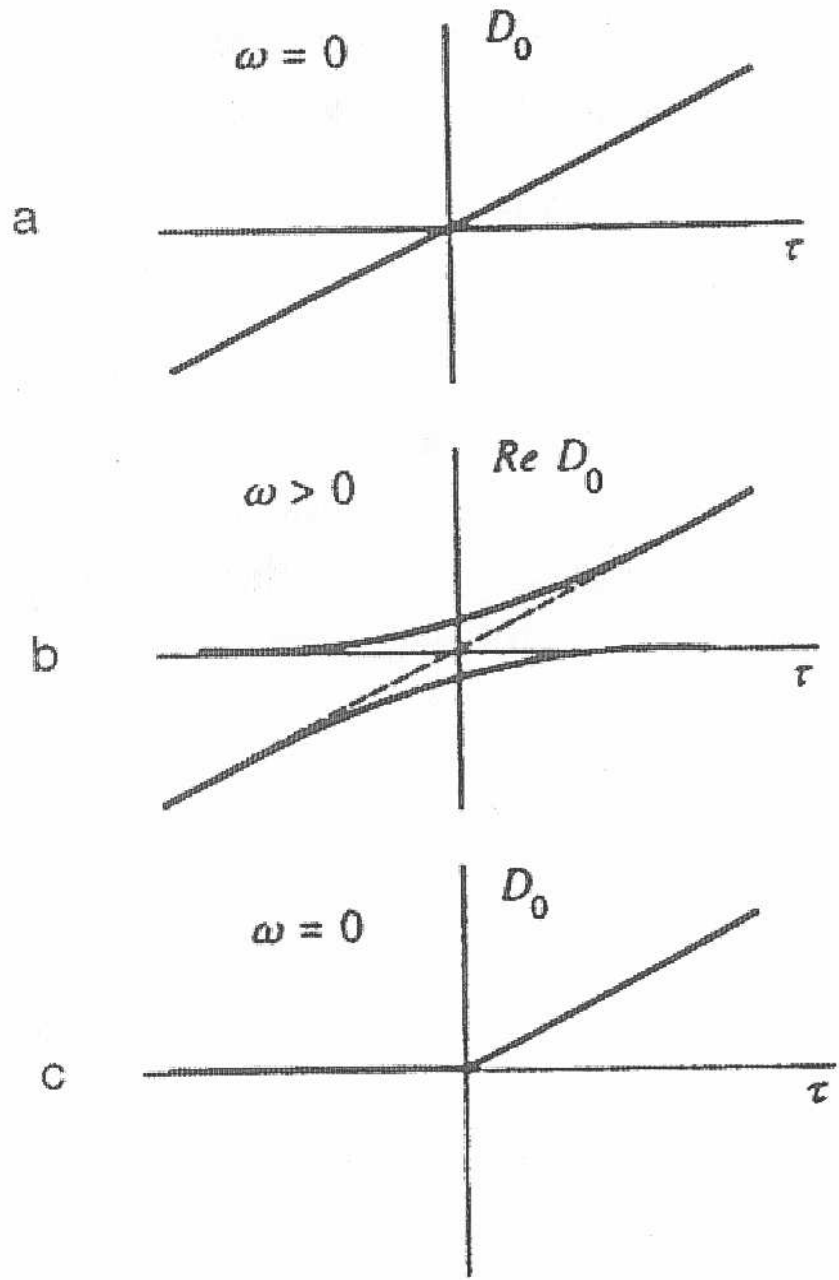}} \caption{
\,\,
Selection of the solutions of the self-consistency equation: a ---
Intersection of the terms for $\omega = 0$; b --- splitting into
physical and unphysical branches for $\omega > 0$; c --- behavior
of $D_0 (\tau)$ for the physical branch at $\omega = 0$. }
\label{fig5}
\end{figure}
For $\omega = 0$
the terms $D_0 = A \tau$ and $D_0 = 0$ are intersected (Fig. 5a); for finite
$\omega$, the degeneracy is removed by the amount $\sim \omega^{1/2 \nu +1}$
(Fig. 5b), and of the two branches, only one satisfies the condition
$\mbox{Re} D(\omega, q) \geq 0$, following from Eqs. (31) and (35) and the
non-negativity of $\langle \rho_E \rho_{E+ \omega} \rangle_{\bf q}$.
Choosing the indicated branch and passing to the limit $\omega \rightarrow
0$, we obtain finiteness of $D_0$ only on one side of the transition --- for
definiteness, for $\tau > 0$ (Fig. 5c).

From Eq. (107) we have for the function $\varphi (q)$
$$
\varphi (q) = {\hat Q}_R^{-1} {\hat P}_\bot \left ( D_0 \phi_0 (q) - \tau
f(q) + \frac{i \omega}{D_0} {\hat Q}_R q^{-2} \right ),\eqno(113)
$$
where ${\hat P}_\bot$ is a projection operator onto the subspace which is
orthogonal to ${\bar \phi}_0 (q)$. Since ${\hat Q}_R \sim 1$ (see below),
we obtain $\varphi (q) \sim \mbox{max} \{ | \tau |, \omega^{1/(2 \nu + 1)}
\}$, which justifies the assumption $\varphi (q) \ll \phi_0 (q)$. For $d >
2$ the integral in Eq. (108) is determined by large values of $\tilde q$
for any regular function $\varphi ({\tilde q})$ and all eigenvalues
$\mu_s^R$ of the operator ${\hat Q}_R$ are found to be of order unity. For
the operator $\hat Q$ from Sec. 5 (which differs from ${\hat Q}_R$ in the
localized phase by the factor $\xi^{-2}$) this means that all $\mu_s$
vanish according to the same law. Therefore the assumption that one mode
predominates is not confirmed by the result and actually the case (b) of
Sec. 6.1 is realized. \\

\begin{center}
{\bf 6.3. Case of several dominant modes}
\end{center}

We seek $D(\omega, q)$ in the form (104), where the choice of the function
$\varphi (q)$ is fixed by the conditions $({\bar \phi}_0, \varphi) = 0, \,
\, C_i = \mbox{const} (\tau)$ as $\tau \rightarrow 0$ (if $\varphi$ is
required to be orthogonal to ${\bar \phi}_1, ..., {\bar \phi}_p$, then the
coefficients $C_i$ are functions of $\tau$, and this leads to
inconveniences in defining the operator ${\hat {\bar L}}_1$ corresponding
to the limit $\omega, \tau \rightarrow 0$ and not depending on $\tau$).
Using instead of $\phi_0$ the "correct" linear combination
$\phi_0 + C_1 \phi_1 + ... + C_p \phi_p$ and repeating the agruments of
Sec. 6.2., we arrive at equations of the type (107) and (108); forming the
scalar product of the first equation with ${\bar \phi}_0$, we arrive at
Eq. (112) with all consequences following from this. Once again, all
eigenvalues of $\hat Q$ vanish according to the same law and the limit
$p \rightarrow \infty$ must be taken in Eq. (104), i.e., all $\phi_s (q)$
must be included in the correct linear combination. Forming the scalar
product of the analog of Eq. (107) with ${\bar \phi}_1, {\bar \phi}_2,
...$, we obtain a system of equations for $C_i$:
$$
D_0 C_i ({\bar \phi}_i, \phi_i) = \tau ({\bar \phi}_i, f) + \mu_i^R ({\bar
\phi}_i, \varphi ) -
$$
$$
- \frac{i \omega}{D_0} ({\bar \phi}_i, {\hat Q}_R q^{-2}), \, \, \, \qquad
\, i = 1, 2, ...,\eqno(114)
$$
where $\mu_i^R \sim 1$. The function $\varphi (q)$ is found to be $\sim
\tau$, and in the limit $\omega \rightarrow 0$, it has a discontinuity at
$\tau = 0$, i.e.,
$$
\varphi (q) = \tau \left \{ \begin{array}{cc}
B_1^M \phi_1 (q) + B_2^M \phi_2 (q) + ..., & \tau > 0 \\
B_1^D \phi_1 (q) + B_2^D \phi_2 (q) + ..., & \tau < 0
\end{array} \right . .\eqno(115)
$$
Substituting the expression (115) into Eq. (114), we obtain in the limit
$\omega \rightarrow 0$ the equations
$$
D_0 C_i ({\bar \phi}_i, \phi_i) = \tau ({\bar \phi}_i, f) + \tau \mu_i^R
B_i^M ({\bar \phi}_i, \phi_i)
$$
$$
0 = \tau ({\bar \phi}_i, f) + \tau \mu_i^R B_i^D ({\bar \phi}_i, \phi_i
) + \xi^{-2} ({\bar \phi}_i, {\hat Q}_R q^{-2})\eqno(116)
$$
for the metallic and dielectric phases, respectively. For any $C_i$ the
equations (116) can be satisfied by appropriate choice of $B_i^M$ and
$B_i^D$, i.e., the coefficients of the correct linear combination are
completely arbitrary. The meaning of this arbitrariness will be explained
in Sec. 8.

Finally, $D (\omega, q)$ near the transition has the form
$$
D (\omega, q) = D_0 {\bar d} (q),
$$
$$
D_0 \sim \left \{ \begin{array}{cc}
\tau, & \tau \gg \omega^{1/(2 \nu +1)} \\
\omega^{1/(2 \nu +1)}, & | \tau | \alt \omega^{1/(2 \nu +1)} \\
(- i \omega) |\tau |^{-2 \nu}, & - \tau \gg \omega^{1/(2 \nu +1)} \end{array}
\right . ,\eqno (116a)
$$
where the function ${\bar d} (q) \equiv d (q)/d(0)$ varies on the scale
$q \sim \Lambda$. This result, obtained for $D (\omega, q)$ defined as in
Eq. (55), is also valid for the observed diffusion coefficient $D_{obs}
(\omega, q)$, since the renormalizations associated with the functions
$B ({\bf q})$ and $C ({\bf q})$ in Eqs. (53) and (54) either contain no
anomalous dispersion or they are small. \\

\begin{center}
{\bf 7. UNIQUENESS OF THE SOLUTION} \\
\end{center}

In this section the self-consistency equation (99) in the localized phase
is investigated assuming for $d (q)$ an expansion of the general form
(64). \\

\begin{center}
{\bf 7.1. Method of supporting points}
\end{center}

In what follows, integrals of the form
$$
I_k = \int \frac{d^d q}{(2 \pi)^d} \frac{q^{2k}}{\xi^{\alpha_0} +
\xi^{\alpha_1} q^2 + ... + \xi^{\alpha_n} q^{2n} + ...}, \qquad \alpha_n
\geq 0 \eqno(117)
$$
are essenrial. The asymptotic expressions of these integrals in the
limit $\xi \rightarrow \infty$ are calculated by the method of "supporting
points". We  choose an appropriate scaling in order to make the
integrals dimensionless, making the substitution $q = \xi^{-b} t$, and
removing the common factor $\xi^a$ from the denominator; as a result,
the indices $\alpha_s$ become $\alpha_s - a - 2sb$. By choosing
appropriate values of $a$ and $b$ the exponents in two terms of
the denominator in Eq. (117) can be made to be zero and the remaining
exponents become negative. Then
$$
I_k = \xi^{-a-b(d+2k)} \int \frac{d^d t}{(2 \pi)^d}
 \frac{t^{2k}}{t^{2s_1} + t^{2s_2} + \sum \limits_{s \neq s_1, s_2}
\xi^{\alpha_s -a-2sb} t^{2s}} \sim \xi^{-a-b(d+2k)}.\eqno(118)
$$
The integer numbers $s_1$ and $s_2$ should satisfy the condition $2s_1 < d +
2k < 2s_2$, guaranteeing that the integral converges after the sum over $s$
is dropped; to avoid uncertainties, we assume that $d$ is noninteger, and
pass to the limit of integer $d$ in the final results. The procedure
described above admits a simple geometric interpretation\footnote{Similar
constructions arise in the investigation of Burgers equation [45].}. Let us
plot a sequence $\alpha_n$  against $n$ (Fig. 6a), mark on the abscissa
\begin{figure}
\centerline{\includegraphics[width=5.1 in]{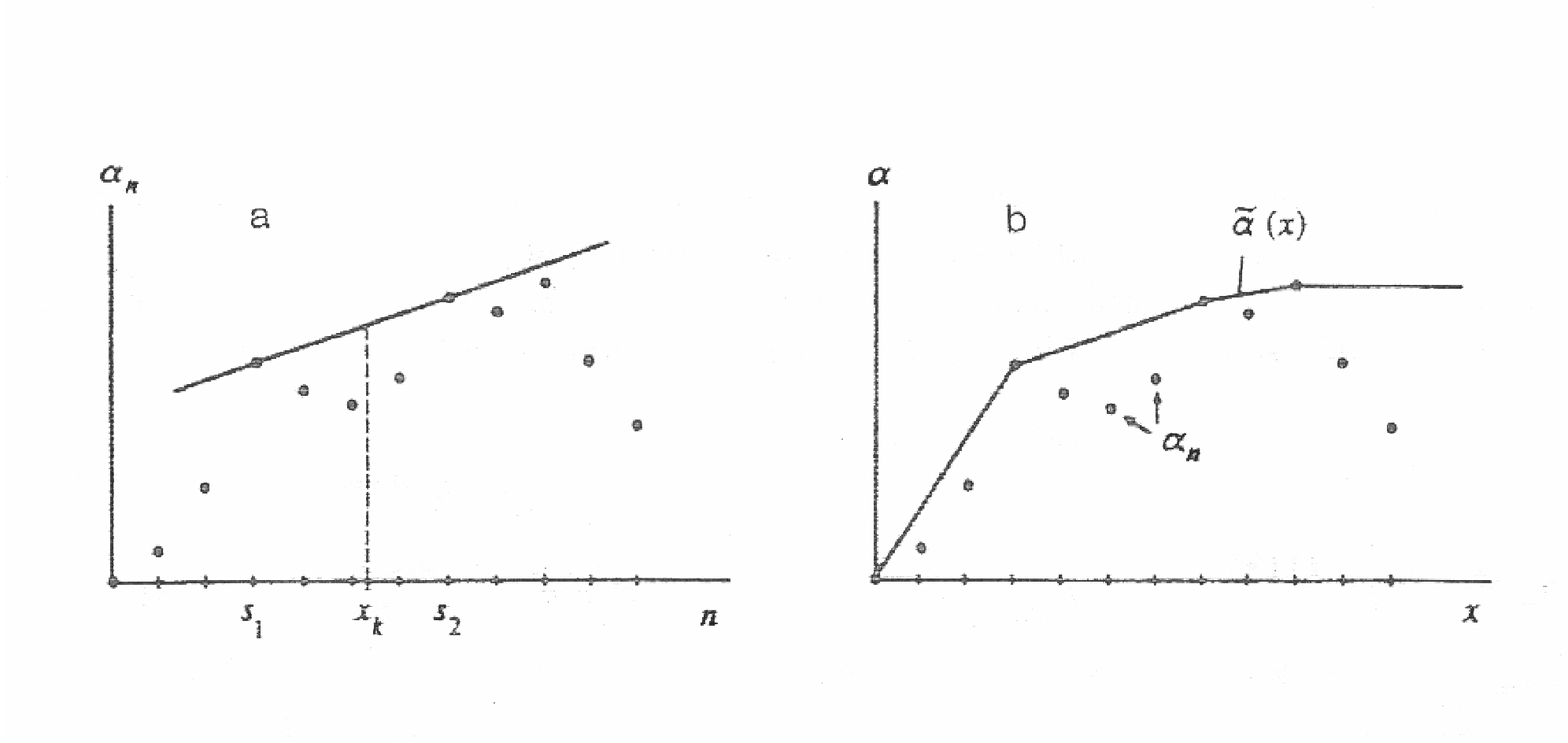}} \caption{
\,\,a --- Construction of the upper
tangent at the point $x_k$ to the
set of points $(n, \alpha_n)$; b --- convex envelope ${\tilde
\alpha} (x)$ for the sequence $\alpha_n$. } \label{fig6}
\end{figure}
axis the point $x_k = (d + 2k)/2$, and construct "an upper tangent" at the
point $x_k$ to the set of points $(n, \alpha_n)$. If we imagine, that the
points are represented by nails, then this construction is made
with the aid of "stick" (solid line in Fig. 6a) and "rope" (dashed line).
The numbers of the points, on which the upper tangent "lies", determine $s_1$
and $s_2$, and its equation $\alpha = a + 2bn$ determines the parameters $a$
and $b$.

Constructing the broken line, consisting of segments of the upper tangents
(Fig. 6b), we obtain a "convex envelope" ${\tilde \alpha} (x)$, in terms of
which the result (118) assumes the form
$$
I_k \sim \xi^{- {\tilde \alpha} (x_k)}, \qquad \, \, \, \, x_k =
\frac{d+2k}{2}.\eqno(119)
$$
By construction the function ${\tilde \alpha} (x)$ is increasing and
convex (in the not strict sense). For a bounded sequence $\alpha_n$ with
maximum at $n = n_0$, ${\tilde \alpha} (x)$ is strictly increasing for $x
< n_0$ and constant for $x > n_0$ (for $x_k > n_0$ the supporting point
$s_2$ lies at infinity). For a strictly increasing and strictly convex
sequence $\alpha_n$, the following inequalities follow from Eq. (119):
$$
I_0 \gg I_1 \gg I_2 \gg ... \gg I_k \gg I_{k+1} \gg ...,\eqno(120)
$$
$$
I_{k_1} I_{k_2} \ll I_{k_1-1} I_{k_2+1} \ll I_{k_1-2} I_{k_2+2} \ll ... \, , \,
\qquad \, \, \, k_1 \leq k_2.\eqno(121)
$$
For an arbitrary sequence $\alpha_n$, some of the strong inequalities are
replaced by weak inequalities. In what follows, for definiteness, we
proceed from the strong inequalities in Eqs. (120) and (121), having in
mind  that the results remain valid in the order of magnitude for the
general case. \\

\begin{center}
{\bf 7.2. Symmetrization of the operator ${\bf {\hat Q}}$}
\end{center}

We set in Eq. (99) $\delta f(q) \equiv \delta \tau f (q)$ and expanding
all functions in series
$$
f (q) = \sum \limits_{k=0}^\infty f_k q^{2k}, \, \, \, \, \qquad \delta d(q) =
\sum \limits_{k=0}^\infty \delta d_k q^{2k}, \, \, \, \, \qquad
B (q, {\tilde q}) = \sum \limits_{k, k^\prime=0}^\infty B_{kk^\prime}
q^{2k} {\tilde q}^{2k^\prime}\eqno(122)
$$
we obtain in the limit $\omega \rightarrow 0$
$$
- \delta \tau f_k = \sum \limits_{k^\prime =0}^\infty B_{kk^\prime} \sum
\limits_{k^{\prime \prime}=0}^\infty I_{k^\prime + k^{\prime \prime}}
\delta d_{k^{\prime \prime}},\eqno(123)
$$
where
$$
I_k = \int \frac{d^dq}{(2 \pi)^d} \frac{q^{2k+2}}{[1 + d(q)
q^2]^2}.\eqno(124)
$$
The matrix ${\hat B} = || B_{kk^\prime} ||$ and the column matrix ${\hat
f} = || f_k ||$ are of general form with elements $\sim 1$. The column
matrix ${\hat {\tilde f}} = {\hat B}^{-1} {\hat f}$ has the same
properties. Multiplying Eq. (123) by ${\hat B}^{-1}$ we obtain
$$
- \delta \tau \left | \begin{array}{c}
{\tilde f}_0 \\
{\tilde f}_1 \\
{\tilde f}_2 \\
\vdots \end{array} \right | = \left | \begin{array}{cccc}
I_0 & I_1 & I_2 & \ldots \\
I_1 & I_2 & I_3 & \ldots \\
I_2 & I_3 & I_4 & \ldots \\
\ldots & \ldots & \ldots & \ldots \end{array} \right | \left |
\begin{array}{c}
\delta d_0 \\
\delta d_1 \\
\delta d_2 \\
\vdots \end{array} \right | ,\eqno(125)
$$
i.e., an equation of the type (99) but with the symmetrized matrix of the
operator $\hat Q$.

Using the expansion (64) for $d (q)$, the integrals (124) acquire the form
(117) with the exponents
$$
\alpha_n = \mbox{max}_{ \{ k_i \} } (\beta_{k_1} + \beta_{k_2})_{k_1 + k_2
= n}\eqno(126)
$$
and an extra $q^2$ in the numerator. The exponents $\beta_n$ are
nonnegative and increase not faster than $2n$ (Sec. 4). This guarantees
the condition $\alpha_n \geq 0$ and make is possible to construct a convex
envelope. \\

\begin{center}
{\bf 7.3. Inversion of the operator ${\bf {\hat Q}}$}
\end{center}

Restricting the upper limit of the summation in Eq. (122) by some finite $n$,
we obtain in Eq. (125) a system of equations of finite order that can be
solved by Cramer's rule. The determinant of the matrix $\hat Q$ in Eq.
(125) consists of all possible products of the form
$$
I_{k_0} I_{k_1 + 1} I_{k_2 + 2} ... I_{k_n + n},\eqno(127)
$$
where $k_0, k_1, ..., k_n$ is a permutation of $0, 1, ..., n$. We separate
in Eq. (125) the pair $I_{k_s + s}I_{k_s^\prime + s^\prime}$ with $s <
s^\prime$. If $k_s > k_{s^\prime}$, then it follows from Eqs. (120) and
(121) that
$$
I_{k_s^\prime + s} I_{k_s + s^\prime} \gg I_{k_s + s} I_{k_{s^\prime} +
s^\prime}\eqno(128)
$$
and the product (127) can be increased by interchanging $k_s$ and
$k_{s^\prime}$, without touching the other $k_i$. It is obvious that in
the maximum product among the products (127), which determines the order
of magnitude of the determinant $Q$, should have $k_0 < k_1 < ... < k_n$,
whence $k_0 = 0, \, \, k_1 = 1, ... , k_n = n$ and therefore
$$
\mbox{det} Q \sim I_0 I_2 I_4 ... I_{2n}.\eqno(129)
$$
The minor $Q_j^i$ of the matrix $Q$, obtained by crossing out the $i$th
row and the $j$th column, consists of all possible products of the form
$$
I_{k_0} I_{k_1 + 1} ... I_{k_{j-1} + (j-1)} I_{k_{j+1} + (j+1)} ...
I_{k_n+n},\eqno(130)
$$
where $k_0, k_1, ..., k_{j-1}, k_{j+1}, ..., k_n$ is a permutation of $0,
1, ..., i-1, i+1, ..., n$. In the maximum product these two sequences are
identical. It is easily verified that
$$
Q_j^0 \ll Q_j^1 \ll ... \ll Q_j^n \sim I_0 I_2 ... I_{2j-2} I_{2j+1} I_{2j+3}
... I_{2n-1}.\eqno(131)
$$
Solving Eq. (125) by Cramer's rule and using Eqs. (129) and (131), we
obtain
$$
\delta d_k \sim \delta \tau \frac{1}{I_{2k}} \frac{I_{2k+1} I_{2k+3} ...
I_{2n-1}}{I_{2k+2} I_{2k+4} ... I_{2n}}, \qquad \, \, \, \, k = 0, 1,
..., n\eqno(132)
$$
and, using Eq. (119), the result can be expressed in terms of the sequence
$\alpha_k$. For a convex sequence $\beta_k$, we have from Eq. (126)
$\alpha_{2m} = 2 \beta_m, \, \, \alpha_{2m+1} = \beta_m + \beta_{m+1}$,
which can be written in the form $\alpha_k = 2 \beta_{k/2}$, if the
sequence $\beta_k$ is additionally defined at half-integral points by the
relation $\beta_{k+1/2} \equiv (\beta_k + \beta_{k+1})/2$. Since the values
of $\beta_{k+1/2}$ lie on the convex envelope ${\tilde \beta} (x)$, for
arbitrary $x$ we obtain
$$
{\tilde \alpha} (x) = 2 {\tilde \beta} (x/2).\eqno(133)
$$
This result remains valid for an arbitrary sequence $\beta_k$. To prove
this, it is necessary to introduce an auxiliary convex sequence
${\bar \beta}_k = {\tilde \beta} (k) \geq \beta_k$ and note that replacing
$\beta_k$ by ${\bar \beta}_k$ does not change the value of the integrals
$I_k$. For convex $\beta_k$ we have
$$
{\tilde \beta} (k + \varphi) = (1 - \varphi) \beta_k + \varphi
\beta_{k+1}, \qquad \qquad \, \, \, \, 0 \leq \varphi \leq 1,\eqno(134)
$$
which makes it possible to switch from the convex envelope directly to the
values of $\beta_k$. In the general case Eq. (134) is correct with
$\beta_k$ replaced by ${\bar \beta}_k$. Setting
$$
d \equiv 4m + 4 \psi, \qquad \qquad m - \mbox{integer}, \, \, \, \, 0 \leq
\psi \leq 1 \eqno(135)
$$
we obtain from Eqs. (132), (119), (133) and (134) in the limit $n
\rightarrow \infty$
$$
\delta d_k \sim \delta \tau \xi^{S(k)}, \qquad
$$
$$
\, \, S(k) =  \left \{
\begin{array}{cc}
(1-2 \psi ) {\bar \beta}_{m+k} + 2 \psi {\bar \beta}_{m+k+1} + {\bar
\beta}_\infty, & \quad 0 \leq \psi \leq \frac{1}{2} \\
{\bar \beta}_{m+k+1} + {\bar \beta}_\infty, & \quad
\frac{1}{2} \leq \psi \leq
1 \end{array} \right . ,\eqno(136)
$$
where the limit ${\bar \beta}_\infty = \lim_{k \rightarrow \infty}
{\bar \beta}_k$ is assumed to be finite in accordance with the
consideration of the next section. \\

\begin{center}
{\bf 7.4. Impossibility of unbounded growth of ${\bf \beta_k}$}
\end{center}

For an unbounded sequence $\beta_k$ the convex envelope ${\tilde \beta}_k$
is strictly increasing and the hierarchy (120) continues to infinity. By
virtue of Eq. (132), this means that $\delta d_k$ diverges as $n
\rightarrow \infty$. To clarify the reasons for the divergence, we note
that the off-diagonal part of the matrix $Q$ in Eq. (125) under the
conditions (120) and (121) can be regarded as a perturbation. Its
eigenvalues in leading order are equal to $I_{2k}$ and they bunch up near
zero in the limit $k \rightarrow \infty$. In the proof of Fredholm's
theorem it is shown [46] that when the expansions (122) are truncated at
the $n$th term, the $(n+1)$ maximum eigenvalues of the operator $\hat Q$
are reproduced; in the limit $n \rightarrow \infty$, arbitrary small
eigenvalues are reproduced and the response of the system to a small
perturbation diverges. This situation occurs not only at the transition
point but also in a neighborhood of the transition point (as long as
$\xi \gg \Lambda^{-1}$); it is unphysical, since the system is unstable
with respect to an infinitely small perturbation of a general form.

This result has important qualitative consequences, since it excludes the
cases corresponding to Fig. 3b and proves the validity of the results
$D (\omega, 0) \sim (-i \omega ) \xi^2$ and $\epsilon (0, 0) \sim \xi^2$
in the localized phase. \\

\begin{center}
{\bf 7.5. Change in ${\bf d (q)}$ as the transition is approached}
\end{center}

Expanding the numerator in Eq. (92) in powers of ${\tilde q}^2$, we obtain
integrals that can be calculated by the method of supporting points and
which are of order $\xi^{- {\tilde \beta}(x_0)}, \xi^{- {\tilde
\beta}(x_1)}$, and so on. We set
$$
{\hat L}_{sing} (\xi) = \frac{{\hat L}_1 (\xi)}{(-i \omega) \xi^{{\tilde
\beta} (x_0)}} = \frac{{\hat {\bar L}}_1 + \xi^{-y_1} {\hat l}_1 +
\xi^{-y_2} {\hat l}_2 + ...}{(-i \omega) \xi^{{\tilde
\beta} (x_0)}},\eqno(137)
$$
where the terms $\xi^{-y_k} {\hat l}_k$ arise from the higher order terms
in the expansion in ${\tilde q}^2$ and from corrections to the main
scaling in the method of supporting points. We have changed the
definition of ${\hat L}_1$ in comparison with (77), in order to make it
possible to separate the main singularity as $\tau \rightarrow 0$ and
introduce the operator ${\hat {\bar L}}_1$, corresponding to the limit
$\omega, \tau \rightarrow 0$.

In Eq. (137) it was assumed that $\beta_k$ are constant. Now, let the
change $\delta \tau$ in the parameter $\tau$ generate the changes $\delta
\xi$ and $\delta \beta_k$ in the quantities $\xi$ and $\beta_k$. Then
$$
{\hat L}_{sing} (\xi + \delta \xi) = \frac{{\hat L}_1 (\xi + \delta
\xi)}{(-i \omega) (\xi + \delta \xi)^{{\tilde \beta} (x_0)}} =
$$
$$
= \frac{{\hat
{\bar L}}_1 + (\xi + \delta \xi)^{{\tilde \beta} (x_0)} (-i) \delta {\hat
L}_1 + (\xi + \delta \xi)^{-y_1} {\hat l}_1}{(-i \omega) (\xi + \delta
\xi)^{{\tilde \beta} (x_0)}},\eqno(138)
$$
where only the term with the minimum index $y_1$ is retained, and $\delta
{\hat L}_1$ is determined by the expression (93) with $\delta d(q)$ of the
form
$$
\delta d (q) = \sum \limits_{k=0}^\infty q^{2k} \xi^{\beta_{k+1}}
\mbox{ln} \xi \delta \beta_{k+1}.\eqno(139)
$$
Using as $\delta {\hat L}_1$ in Eq. (97) the quantity ${\hat
L}_1 (\xi + \delta \xi) - {\hat L}_1 (\xi)$ we obtain instead of Eq. (99)
$$
\delta D (\omega, q) = \delta \tau f (q) + \xi^{-y_1-1} \delta \xi R(q) +
\xi^{{\tilde \beta} (x_0)} {\hat Q} \delta d(q),\eqno(140)
$$
where $\delta D (\omega, q) \rightarrow 0$ as $\omega \rightarrow 0$. In
the case of exact scaling, when $\delta d (q) \equiv 0$, the first two
terms on the right-hand side cancel one another, whence $y_1 = 1/ \nu$. In
the general case, they are of the same order of magnitude. Inverting the
operator $\hat Q$ according to Sec. 7.3 and comparing with Eq. (139), we
have
$$
\delta d_k \sim \delta \tau \xi^{- {\tilde \beta} (x_0) + S(k)} \sim
\xi^{\beta_{k+1}} \mbox{ln} \xi \delta \beta_{k+1},\eqno(141)
$$
and, consequently
$$
\delta \beta_k \sim \frac{\delta \tau}{\mbox{ln} \xi}
\xi^{\gamma_k},\eqno(142)
$$
where it is convenient to write $\gamma_k$ in the form
$$
\gamma_k =
 \left \{ \begin{array}{cc}
({\bar \beta}_k - \beta_k) + ({\bar \beta}_\infty - {\tilde \beta} (x_0))
+ ({\bar \beta}_{m+k} - {\bar \beta}_k) &
\quad (\frac{1}{2} \leq \psi \leq 1) \\
{  }&{ }\\
({\bar \beta}_k - \beta_k) + ({\bar \beta}_\infty - {\tilde \beta} (x_0))
+ (1 - 2 \psi) ({\bar \beta}_{m+k-1} - {\bar \beta}_k) + & \quad (0 \leq \psi
\leq \frac{1}{2}, \\
+ 2 \psi ({\bar \beta}_{m+k} - {\bar \beta}_k) & \quad m \geq 1) \\
{  }&{ }\\
({\bar \beta}_k - \beta_k) + ({\bar \beta}_\infty - {\bar \beta}_k)
+ (1 - 2 \psi) ({\bar \beta}_{k-1} - {\bar \beta}_0) + & \quad (0 \leq \psi
\leq \frac{1}{2}, \\
+ 2 \psi ({\bar \beta}_k - {\bar \beta}_1) & \quad m = 0)
\end{array} \right . .\eqno(143)
$$
All combinations in parentheses are non-negative and $\gamma_k \geq 0$.
For fixed $\xi$ the definition of the exponents $\beta_k$ in Eq. (64) is
not unique: the coefficient of $q^{2k}$ can be written as $C_k \xi^{\beta
k}$ and a change in $\beta_k$ is equivalent to a change in $C_k$. The
specific configuration of the exponents $\beta_k$ make sense only if it
remains unchanged when $\tau$ changes. According to Eq. (142), for
$\gamma_k > 0$ a large change in the exponents occurs.
Only if $\gamma_k \equiv 0$, the changes $\delta
\beta_k \sim \delta \tau / \mbox{ln} \xi$ can be included in the changes
of $C_k$. The condition
$\gamma_k \equiv 0$ requires that all combinations in parentheses of Eq.
(142) vanish and it fixes the only configuration of exponents, which is
different for $d > 2$ and $d < 2$:
$$
\beta_1 = \beta_2 = \beta_3 = ... \, , \qquad \, \, \, \, d > 2,
$$
$$
\beta_0 = \beta_1 = \beta_2 = ... \, , \qquad \, \, \, \, d < 2.\eqno(144)
$$
By definition, $\beta_0 = 0$ and for $d < 2$ all exponents are equal to zero.
This means that $d (q)$ does not diverge and the localized phase remains
for all $\tau$ [21]. For $d > 2$, all indices can be made equal to 2, in
accordance with the requirement $\beta_1 = 2$ (Fig. 3a), by redefining $\xi$.
All eigenvalues of $\hat Q$ vary according to the same law and we return
to be case (b) of Sec. 6.1. \\

\begin{center}
{\bf 8. CHANGE IN SYMMETRY AT THE ANDERSON TRANSITION} \\
\end{center}

A change in ${\hat L}_{reg}$ gives rise to a rotation of the subspace
$M_\infty$ of the operator ${\hat L}_{sing}$. This is analogous to a
rotation of the magnetization vector $\bf M$ accompanying a change in the
magnetic field $\bf H$ in a ferromagnet. This analogy is formalized in the
form of Table 1. We shall give some explanations\footnote{A similar, but
not identical, analogy was discussed in Ref. [41].}.

The operators ${\hat L}_{reg}$ and ${\hat L}_{sing}$ have many degrees of
freedom, many of which do not appear in the self-consistency equation. The
important degrees of freedom are determined by the functions $f (q)$ and
${\bar d} (q)$ [see Eq. (116a)], whose expansion coefficients
$$
f(q) = 1 + f_1 q^2 + ... + f_n q^{2n} + ...,
$$
$$
{\bar d} (q) = 1 + d_1 q^2 + ... + d_nq^{2n} + ...\eqno(145)
$$
can be regarded as components of the unit vectors ${\hat {\bf H}}$ and
${\hat {\bf M}}$. In the localized phase, small changes in them are related
with each other by the operator $\hat Q$, whose inverse is analogous to the
magnetic susceptibility tensor $\chi_{ij}$.

The finiteness of the frequency $\omega$ smears the transition, similarly
to the finiteness of the magnetic field in a ferromagnet. In the localized
phase $D_0 \sim \omega$ and in the metallic phase $D_0 = \mbox{const}
(\omega)$, which is analogous to the appearance of spontaneous
magnetization, i.e., the quantities $D_0$ and $\omega$ are analogous to
$| {\bf M} |$ and $| {\bf H} |$. In view of the qualitative character of
the analogy, this identification is not unique. For example, any monotonic
function $F (| {\bf M} |)$, equal to zero for $| {\bf M} | = 0$, can be
taken as the analog of $D_0$; as a function of the magnetization itself,
it has the form $F (M^2)$, since a scalar must be formed from a vector.
Finally, for small $M^2$, it can be expanded in a series, and we obtain
an analogy of $D_0$ to $M^2$. Similarly, $H^2$ is the natural analog for
$\omega$.

\begin{center}
{\rm TABLE 1.} {\rm Analogy between a ferromagnet and a disordered system.}
\vspace{3mm}
\begin{tabular}{||c|c||} \hline
Ferromagnet & Disordered system \\ \hline
Orientation of the magnetic field $\bf H$: & Operator ${\hat L}_{reg}$: \\
Components of the unit vector & Coefficients $f_n$ \\
Orientation of the magnetization $\bf M$: & Space $M_\infty$ of the
operator ${\hat L}_{sing}$ \\
Components of the unit vector & Coefficients $d_n$ \\
Squared modulus of the field $H^2$ & Frequency $\omega$ \\
Squared modulus of the & Diffusion coefficient $D_0$ \\
magnetization $M^2$ & \\
Magnetic susceptibility tensor $\chi_{ij}$ & Operator ${\hat Q}^{-1}$ \\
Paramagnetic phase & Localized phase \\
Ferromagnetic phase & Metallic phase \\
Curie point & Point of the Anderson transition \\
$T - T_c$ & Distance to the transition $\tau$ \\ \hline
\end{tabular}
\end{center}

In the analogy found, it is important that (a) the number of components of
the vector $\bf M$ is infinite, since the number of expansion coefficients
$d_n$ is infinite, and (b) the ferromagnet is isotropic. The latter
property is obvious from the fact that all eigenvalues of the
"susceptibility tensor" ${\hat Q}^{-1}$ diverge at the transition point
according to the same law and for small changes in $d_n$ and $f_n$ they
can be made equal by a $\tau$-independent linear transformation. The
analog of collinearity of $\bf M$ and $\bf H$ in an isotropic ferromagnet
exists with the following stipulation. In the case of
the Anderson transition the "vector $ {\bf H}$" and the "vector $ {\bf
M} $" lie in different subspaces and there is no natural method for
establishing the mutual orientation of these subspaces. For this reason,
for a fixed function $f(q)$ the choice of $d(q)$ is arbitrary (Sec. 6.3),
in accordance with the arbitrariness in the choice of bases in the two
subspaces. For the special choice $B_i^M \equiv 0$ in Eq. (115) we have
$f(q) \equiv {\bar d} (q)$, which corresponds to the choice of the
"correct" mutual orientation of the bases.

The model of an isotropic ferromagnet with the number of components $n
\rightarrow \infty$ is well known in the theory of phase transitions and
is the basis for the $1/n$ expansion [1]. Its critical exponents are known
exactly. Specifically, for the exponents of the magnetization $M
\sim \tau^\beta$ and of the correlation length  $\xi \sim \tau^{-
\nu}$ we have
$$
\beta = 1/2\,\,, \quad d > 2\,\,;
 \qquad \qquad  \nu = \left \{ \begin{array}{cc}
{\displaystyle \frac{1}{d-2}}\quad , & 2 < d < 4 \\ {   } & {  } \\
{\displaystyle \frac{1}{2}} \quad, & d > 4
\end{array} \right . \quad.
\eqno(146)
$$
which, since $s = 2 \beta$, corresponds exactly to Eq. (18). \\

\begin{center}
{\bf 9. CONCLUSIONS} \\
\end{center}

The approach to the theory of localization based on the formalism of
$\sigma$-models [22, 47, 48] is currently considered to be the most
rigorous approach. However, its rigor should not be overestimated. First,
the degree to which the approximations employed in the derivation of the
$\sigma$-models retain exact time-reversal invariance  of the initial
disordered system and the satisfaction of the Ward identity (13),
was never studied. But these properties are vitally important
  for reproducing the pole structure of $U_{{\bf
kk}^\prime} ({\bf q})$. Second, to take into account the
spatial dispersion of $D (\omega, {\bf q})$, it is necessary to introduce
into the Lagrangian of the $\sigma$-model additional gradient vertices,
which grow anomalously at the initial stage of the renormalization group
transformations [49]. The analog of such growth can be obtained from Eqs.
(142) and (143), assuming for the initial configuration of exponents
$\beta_1 = 2, \, \, \beta_k = 0$ ($k \geq 2$):
$$
\frac{\partial \beta_k}{\partial \tau} \sim \frac{1}{\mbox{ln} \xi} \xi^{2
- \beta_k}, \qquad \, \, \, \, \, \, k = 1, 2, ... .\eqno(147)
$$
Growth of $\beta_k$ with $k \geq 2$ indicates intensification of spatial
dispersion of $D (\omega, {\bf q})$ as the transition is approached.
In the language of the magnetic analogy (Sec. 8), it corresponds to a
transformation of a uniaxial ferromagnet into an isotropic
ferromagnet\footnote{A detailed investigation of the evolution of $\beta_k$
requires a knowledge of the proportionality coefficients in Eq. (64) and
(142).}. One can speculate, that the renormalization group transformations
can transfer analogously the zero-component $\sigma$-model into
an infinite component model. These difficulties apparently are not important
in low orders in $\epsilon = d - 2$, since for small $\epsilon$ the Anderson
transition falls in the region of weak disorder, for which the derivation of
the $\sigma$-model is indeed substantiated.

We now discuss the possible reasons for the disagreement of Eq. (18) with
the result of Ref. 28 for the exponent $s$. The result of Ref. 28 for the
permittivity $\epsilon (0, 0) \sim \xi$ corresponds to the case $\zeta =
1$ of the Sec. 4. This is also indicated by the expression given in Ref.
28 for the function $A ({\bf r})$ from Eq. (66). For $\zeta = 1$ the
exponents $\beta_k$ increase linearly with $k$, and in accordance
with Sec. 7.4, the instability
with respect to an infinitesimal perturbation of a general form arises.
If the results of Ref. 28 correspond to the exact solution of some
idealized model, then this model is unphysical. More
likely, the approximations employed in the derivation of the
$\sigma$-model and the selection of diagrams in Ref. 28 destroy the pole
structure of $U_{{\bf kk}^\prime} ({\bf q})$. The fact,
that results  are the same for models with and without the time-reversal
invariance, looks suspicious in this sense.
Finally, in the derivation of the $\sigma$-model for a large $d$ it is
necessary to introduce an artificial construction  of weakly
coupled granules. For this construction, because of the presence
of artificial small parameters, the critical region can narrow anomalously
and, as a result of the approximations, contract into a point. The results of
Ref.  28 could correspond to some intermediate asymptotic behavior.

There are some objections to the arguments of Efetov [43] that the
diagrammatic approach in principle does not "feel" the noncompactness,
which in his opinion determines the main difference between the theory of
disordered systems and the theory of phase transitions. One can agree with
the last assertion: Noncompactness is a consequence of the infinitesmall
additions $\pm i \delta$, determining the type of Green's function,
which lead to nonperturbative contributions giving rise to the difference
between the two indicated theories [17]. However, the nonperturbative
contributions can be obtained from the diagrammatic technique [16]. The
additions $\pm i \delta$ play an important role in the separation of the
diffusion poles, since as a result of these additions, the integration
contour in Eq. (58) is confined between the poles of two Green's functions.

I thank A. F. Andreev, A. I. Larkin, M. V. Sadovskii, and M. V. Feigelman
for a discussion. I am grateful to the participants of the seminars at
P.L.Kapitza Institute for Physical Problems and P.N.Lebedev Physics Institute
of the Academy of Sciences for their interest in this work.

Financial support for this work was provided by the International Science
Foundation (grant MOH000) and the Russian Fund for Fundamental Research
(project 93-02-20). \\

\begin{center}
{\bf APPENDIX} \\
{\bf Expansion of the self-consistency equation in powers of $\bf q$} \\
\end{center}

If we have for the operator $\hat L$
$$
{\hat L} ({\bf q}) \psi_{\bf k} \equiv \frac{1}{N} \sum \limits_{{\bf
k}^\prime} L_{{\bf kk}^\prime} ({\bf q}) \psi_{{\bf k}^\prime},
$$
$$
L_{{\bf kk}^\prime} ({\bf q}) = L_{{\bf k}^\prime {\bf k}} ({\bf q}) =
L_{- {\bf k}, - {\bf k}^\prime} (- {\bf q})\eqno(A1)
$$
then it is easy to prove that: (a) the eigenvalues of $\hat L$ are even as
a function of $\bf q$, $\lambda_s ({\bf q}) = \lambda_s (- {\bf q})$; (b)
the eigenfunctions $e_{\bf k}^{(s)} ({\bf q})$ can be chosen so that
$e_{\bf k}^{(s)} ({\bf q}) = e_{- {\bf k}}^{(s)} (- {\bf q})$; and, (c) if
there are several operators of the type (A1), then the matrix elements of
one operator with respect to the eigenfunctions of the other operator are
even functions of $\bf q$. The operators ${\hat L}, \, \, {\hat L}_{reg}$,
and ${\hat L}_1$ are of the form (A1) and, by virtue of (à--c), the
right-hand side of Eq. (97) is even in $\bf q$ and can be expanded in
powers of $q^2$.

Since the operator ${\hat L}_1$ in Sec. 5.1 is independent of $\omega$, we
obtain that in Eq. (86) $O (\omega) \equiv 0$ and the definitions (75) and
(76) are equivalent to the following definitions:
$$
{\hat L}_{reg} \psi_{\bf k} = (\epsilon_{{\bf k}+{\bf q}/2} - \epsilon_{{\bf
k} - {\bf q}/2}) \psi_{\bf k} + \frac{1}{N} \sum \limits_{{\bf
k}^\prime} U_{{\bf kk}^\prime}^{reg} ({\bf q})
 [\Delta G_{{\bf k}^\prime}
({\bf q}) \psi_{\bf k} - \sqrt{\Delta G_{\bf k} ({\bf q}) \Delta G_{{\bf
k}^\prime} ({\bf q})} \psi_{{\bf k}^\prime} ],\eqno(A2)
$$
$$
{\hat L}_{sing} \psi_{\bf k} = \frac{1}{N} \sum \limits_{{\bf k}^\prime}
U_{{\bf kk}^\prime}^{sing} ({\bf q}) [\Delta G_{{\bf k}^\prime} ({\bf q})
\psi_{\bf k} -
 \sqrt{\Delta G_{\bf k} ({\bf q}) \Delta G_{{\bf k}^\prime}
({\bf q})} \psi_{{\bf k}^\prime} ].\eqno(A3)
$$
For the definition (A2) we have, on account of Eq. (39),
$$
{\hat L}_{reg} |e_0 \rangle = {\hat L}_{reg} \{ \mbox{const} \sqrt{\Delta
G_{\bf k} ({\bf q})} + O ({\bf q}) \} =
$$
$$
=\mbox{const} (\epsilon_{{\bf k} +
{\bf q}/2} -
 \epsilon_{{\bf k} - {\bf q}/2}) \sqrt{\Delta G_{\bf k} ({\bf
q})} + O ({\bf q}) = O ({\bf q})
$$
and the contribution $O (q^0)$ is absent in each term on the
right-hand side of Eq. (97). \\

\end{document}